\documentclass{article}
\usepackage{graphicx}

\usepackage{amsmath}
\usepackage{natbib}
\usepackage{lscape}
\usepackage{xcolor}
\usepackage{mathtools}
\usepackage{hyperref}
\usepackage{lineno}

\usepackage{authblk}

\newcommand{\mcom}{\;\;\textrm{,}}
\newcommand{\mdot}{\;\;\textrm{.}}

\newcommand{\bpp}{b^{(P)}}
\newcommand{\btp}{b^{(T)}}
\newcommand{\jpp}{i^{(P)}}
\newcommand{\jtp}{i^{(T)}}

\newcommand{\refp}[1]{(\ref{#1})}

\newcommand{\Bv}{\mbox{$\mathbf B$}}
\newcommand{\Uv}{\mbox{$\mathbf U$}}
\newcommand{\Ev}{\mbox{$\mathbf E$}}
\newcommand{\Jv}{\mbox{$\mathbf J$}}
\newcommand{\jv}{\mbox{$\mathbf j$}}
\newcommand{\Lv}{\mbox{$\mathbf L$}}
\newcommand{\I}{\mbox{$\mathcal{J}$}}
\newcommand{\Iv}{\mbox{$\boldsymbol{\mathcal{J}}$}}
\newcommand{\Dv}{\mbox{$\boldsymbol{\mathcal{B}}$}}

\newcommand{\U}{\mbox{U}}
\newcommand{\B}{\mbox{B}}
\newcommand{\J}{\mbox{J}}

\newcommand{\be}{\begin{equation}}
\newcommand{\ee}{\end{equation}}
\newcommand{\eep}{\;\;.\end{equation}}
\newcommand{\eec}{\;\;,\end{equation}}
\newcommand{\bea}{\begin{eqnarray}}
\newcommand{\eea}{\end{eqnarray}}
\newcommand{\bel}[1]{\begin{equation}\label{#1}}
\newcommand{\beal}[1]{\begin{eqnarray}\label{#1}}
\newcommand{\tp}[1]{\mbox{$\times10^{#1}$}}
\newcommand{\curl}{\nabla\times}
\newcommand{\uvr}{\hat{\mathbf r}}

\newcommand{\uvz}{\hat{\mathbf z}}

\newcommand{\Ro}{\mbox{Ro}}

\newcommand{\E}{\mbox{E}}
\newcommand{\Ek}{\mbox{E}}

\newcommand{\Pra}{\mbox{Pr}}
\newcommand{\Pm}{\mbox{Pm}}
\newcommand{\Rm}{\mbox{Rm}}

\newcommand{\RmL}{\mbox{Rm$^{(1)}$}}
\newcommand{\RmLS}{\mbox{Rm$^{(2)}$}}
\newcommand{\SDCR}{SDCR}

\newcommand{\Ra}{\mbox{Ra}}

\newcommand{\Di}{\mbox{Di}}

\newcommand{\F}{{\mathbf F}}

\newcommand{\sigmab}{\widetilde{\sigma}}
\newcommand{\lambdab}{\widetilde{\lambda}}
\newcommand{\rhob}{\widetilde{\rho}}
\newcommand{\Tb}{\widetilde{T}}

\newcommand{\gb}{\widetilde{g}}

\newcommand{\Figref}[1]{Fig.~\ref{#1}}
\newcommand{\figref}[1]{fig.~\ref{#1}}

\newcommand{\eqnref}[1]{eqn.~(\ref{#1})}

\newcommand{\Secref}[1]{Sect. \ref{#1}}
\newcommand{\secref}[1]{sect. \ref{#1}}

\newcommand{\tabref}[1]{tab. \ref{#1}}

\title{Dynamo Action in the Steeply Decaying Conductivity Region of 
Jupiter-like Dynamo Models}

\author[1]{J. Wicht}
\author[2]{T. Gastine}
\author[3]{L. Duarte} 

\affil[1]{Max Planck Institute for Solar System Research, 
Justus-von-Liebig-Weg 3, 37077 G\"ottingen, Germany}
\affil[2]{College of Engineering, Mathematics and Physical Sciences,  University of Exeter, 
Physics building,
Stocker Road,
Exeter, EX4 4QL, 
United Kingdom}
\affil[3]{IPGP, 
Institution for Higher Education and Research 1, rue Jussieu, 
75238 Paris cedex 05, France}

\begin{document}

\maketitle

%\correspondingauthor{J. Wicht}{wicht@mps.mpg.de}

\begin{abstract}
The Juno mission is delivering spectacular data of Jupiter's magnetic 
field, while the gravity measurements finally allow constraining  
the depth of the winds observed at cloud level.
However, to which degree the zonal winds contribute to the 
planet's dynamo action remains an open question. 
Here we explore numerical dynamo simulations 
that include an Jupiter-like electrical conductivity profile and 
successfully model the planet's large scale field. 
We concentrate on analyzing the dynamo action in the Steeply
Decaying Conductivity Region (\SDCR) where the high conductivity 
in the metallic Hydrogen region drops to the much lower values 
caused by ionization effects in the very outer envelope of the planet. 
Our simulations show that the dynamo action in the \SDCR\ is 
strongly ruled by diffusive effects and therefore quasi stationary. 
The locally induced magnetic field is dominated by the horizontal toroidal
field, while the locally induced currents flow mainly 
in the latitudinal direction.
The simple dynamics can be exploited to yield estimates of 
surprisingly high quality for both the induced field and the 
electric currents in the \SDCR.
These could be potentially be exploited to predict the dynamo action 
of the zonal winds in Jupiter's \SDCR\ but also in other planets. 
\end{abstract}

%Main text:
        \section{Introduction}
\label{Introduction}

Jupiter's electrical conductivity profile likely 
has a strong impact on the interior dynamics. 
{\it Ab initio} calculations \citep{French2012} show that the conductivity 
first increases with depth at a super-exponential rate 
due to hydrogen ionization. At about $0.90\,r_J$, however, hydrogen undergoes a phase transition from the molecular to 
a metallic state and the conductivity increases  
much more smoothly with depth (see \figref{fig:DensSigma}). 
%In this paper we discuss the possible dynamo action in the outer 
%Steeply Decaying Conductivity Region \SDCR. 

It is commonly thought, that Jupiter's dynamo operates 
in the deeper metallic region, while the fierce  
zonal wind system observed on the surface is limited to the molecular outer shell. However, recent numerical simulations show that 
the interaction between both regions yields 
complex and interesting dynamics. % at least partly harbored by the \SDCR. 
For example, the zonal winds may drive a secondary 
dynamo where they reach down to significant enough 
electrical conductivities. Possible spotting feature of such 
a process would are banded structures and large scale 
spots at low to mid latitudes \citep{Gastine2014,Duarte2018}, very similar to those observed recently by NASA's
Juno mission \citep{Connerney2018}. 

One of the main Juno objectives is to determine the depth of the fierce zonal winds observed 
on the planet's surface.  
Since the winds dynamics is tied 
to density variations \citep{Kaspi2018}, the gravity
signal can help to constrain their depth. 
A recent analysis by \citet{Kaspi2018}, based on the 
equatorially antisymmetric gravity moments $J_3$ to $J_9$ measured
by Juno mission, concludes that the wind speed must be 
significantly reduced at a about $3000\,$km depth, 
which corresponds to about $0.96\,r_J$.  

Older estimates of the wind depth rely on magnetic effects. 
\citet{Liu2008} concluded that 
the Ohmic heat produced by the zonal wind induced 
electric currents would exceed the heating emitted 
from the planets interior, should the winds reach deeper than 
$0.96\,r_J$ with undiminished speed. 
\citet{Ridley2016} argue that the variation of the large 
scale field deduced from pre-Juno measurements is so small
that the zonal winds cannot contribute. 
The winds are thus unlikely to penetrate deeper than 
about $0.96\,r_J$ where the magnetic Reynolds number  
exceeds unity \citep{Cao2017}.

Another hint on the depth of the zonal winds comes from the width 
of the prograde equatorial jet. Numerical simulations suggest that it is determined by 
the depth of the spherical shell the winds live in. 
The observed with is consistent with a lower boundary 
at about $0.95\,r_J$ \citep{Gastine2014a}. 

Classically, the magnetic Reynolds number attempts to quantify the 
ratio of magnetic induction to Ohmic dissipation in the 
dynamo equation
\bel{eq:Dynamo}
\frac{\partial\Bv}{\partial t} = \curl\left(\Uv\times\Bv\right) - \curl\left(\lambda\curl\Bv\right) 
\eec
where $\lambda=1/(\mu\sigma)$ is the magnetic 
diffusivity and $\sigma$ the electrical conductivity. 
The Ohmic dissipation term can be separated into two parts, 
\bel{eq:diff}
\curl\left(\lambda\curl\Bv\right) = 
- \lambda\,\nabla^2\Bv + \frac{\lambda}{d_\lambda}\,\uvr\times\left(\curl\Bv\right)
\eec
where the second depends on the magnetic diffusivity 
scale height $d_\lambda=\lambda/(\partial \lambda/\partial r)$. 
The classical magnetic Reynolds number 
\bel{eq:Rm}
\Rm=\frac{\U d}{\lambda}
\eec
ignores the second term that would dominate where 
$\sigma$ decreases rapidly in Jupiter's outer envelope. 
Here $d$ is a reference length scale and $\U$ typical flow velocity and $d$ a reference 
length scale, 
\citet{Liu2008} argue that the 
appropriate definition for this particular 
Steeply Decaying Conductivity Region (\SDCR)
should include the diffusivity scale height: 
\bel{eq:RmL}
\Rm^{(1)}=\frac{\U d_\lambda}{\lambda}
\eep
In Jupiter, the \SDCR\ roughly coincides with the 
molecular outer envelope where ionization effects determine 
the electrical conductivity. 

For a selfconsistent dynamo to operate, the overall 
field production has to overcome diffusion. The mean global 
magnetic Reynolds number for the whole dynamo region should thus 
exceed a critical value $\Rm_c=1$. However, since the 
definition of $\Rm$ ignores many complexities of the 
dynamo process, the exact critical value is always higher 
and can only be determined by experiments. 
\citet{Christensen2006} explore a suite of Boussinesq 
dynamo simulations with rigid flow boundary conditions and 
constant conductivity and find selfconsistent dynamo action 
when the rms magnetic Reynolds number exceeds 
$\Rm_c\approx 50$.

The radius $r_D$ where the magnetic Reynolds number exceeds 
$\Rm_c$ indicates 
the 'top of the dynamo region' that could potentially 
host selfconsistent dynamo action. \citet{Gastine2014} 
and \citet{Duarte2018} use $\Rm_c\approx 50$ in combination 
with the classical $\Rm$ definition \refp{eq:Rm} 
and show that Jupiter-like 
magnetic field are found when $r_D>0.9\,r_J$. 
Extrapolating their findings to Jupiter conditions  
suggest $r_D\approx 0.95\,r_J$. 

There are several reasons, however, why these consideration
are problematic. For one, self-consistent dynamo 
action and the related critical magnetic Reynolds number are
defined in a mean sense for the 'whole' dynamo region. 
The local dynamo mechanism at any given radius, however, also 
includes the modification of the field produced elsewhere. 
Another fundamental problem is that the magnetic Reynolds number
actually ceases to provide a decent proxy for the 
ratio of induction to diffusion in the \SDCR. 
The reason is that the magnetic field approaches a 
diffusion-less potential field as the conductivity decays. 

A more useful definition of the 'top of the dynamo region'
would be the depth where the locally induced field becomes
a significant fraction of the total field.  
\citet{Cao2017} suggest that $\RmL$ takes on a different role in the 
\SDCR\  and may 
allow to quantify the ratio of locally induced field  
$\hat{\Bv}$ to the background field $\tilde{\Bv}$. 
Using a simplified mean field dynamo model, they 
conclude that $\hat{\Bv}$ could reach $1\,$\% of the 
background field in the \SDCR. 

The different estimates put the depth of the zonal winds in 
the \SDCR\  beyond $0.9\,r_J$. 
In this article we closely analyze the 
dynamo action in the \SDCR\ for two different 
numerical dynamo simulations that both yield 
Jupiter-like large scale magnetic fields. 
After presenting the numerical dynamo model 
in \secref{sec:Model}, we analyze the simulation results 
in \secref{sec:Results}. 
\Secref{sec:Estimates} is then devoted to deriving 
estimates for the locally induced electric current and 
field that could be applied to planets. 
The paper closes with a discussion and conclusions 
in \secref{sec:Discussion}.
\section{Numerical Dynamo Model}
\label{sec:Model}

\subsection{Fundamental Equations}

The numerical simulations were performed with the MHD-code 
MagIC that is freely available on GitHub 
(https://github.com/magic-sph/magic). 
MagIC solves for small disturbances around an 
adiabatic, hydrostatic background state that only depends 
on radius. Using the anelastic approximation for the 
Navier-Stokes equation allows incorporating 
the background density gradient $\rhob(r)$ while filtering 
out fast sound waves.

\begin{figure}[h!]
\begin{center}
{\centering
      \includegraphics[width=0.7\textwidth]{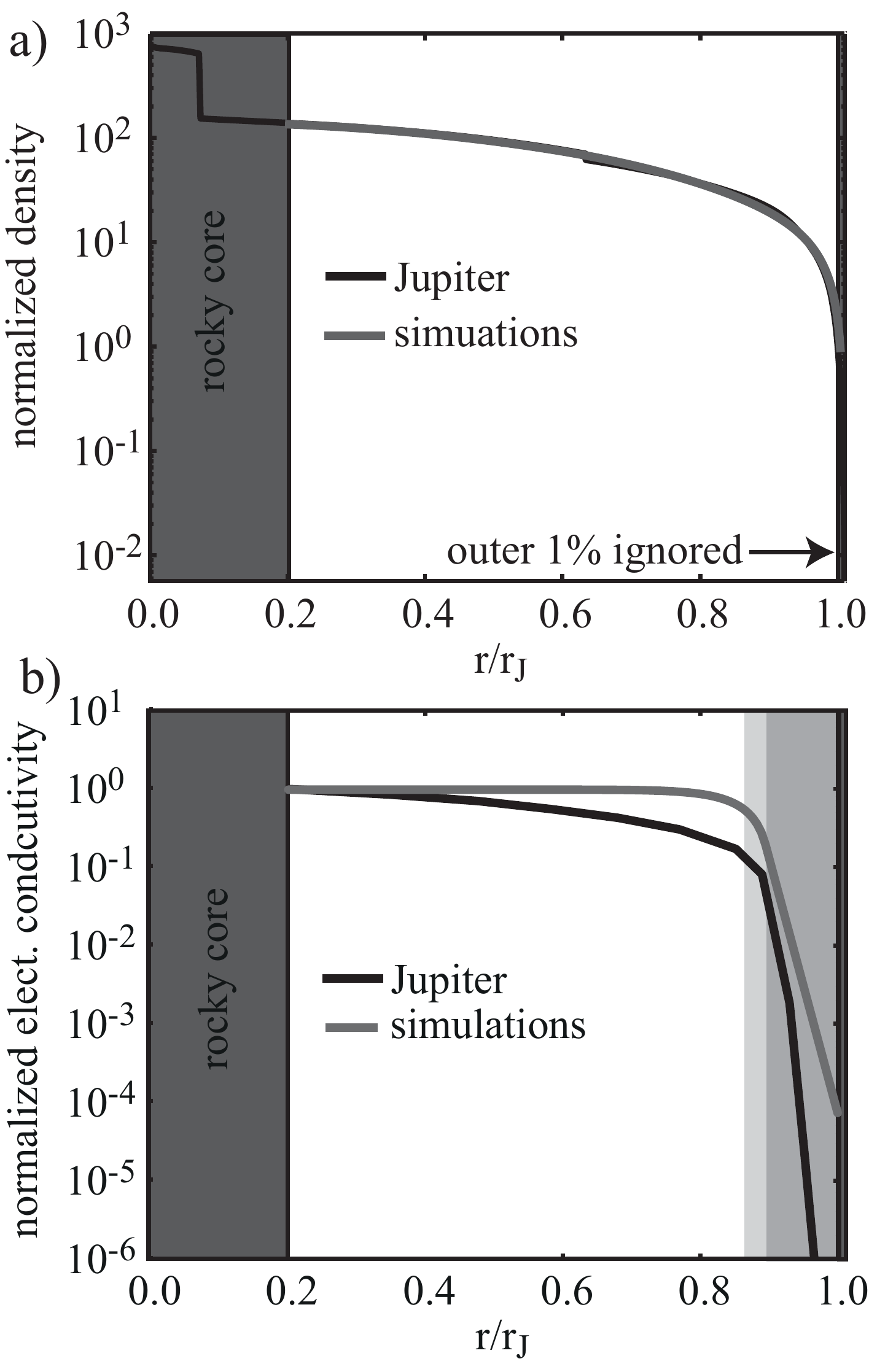}
}
\caption{Comparison of the Jupiter interior model by \citep{French2012} ({\it black lines})
with the model used for the numerical simulations (grey). 
{\it Panel a)} and {\it b)} show the normalized background density 
and electrical conductivity, respectively. Dark grey 
background colors mark the outer $1$\% that have been neglected 
in the simulations and the assumed solid inner core. 
Medium gray highlights the Steeply Decaying 
Conductivity Region (\SDCR) of the model conductivity profile
while Light grey marks the low conductivity region  
where the conductivity is smaller than $50$\% of the inner boundary
reference value.
\label{fig:DensSigma}}
\end{center}
\end{figure}

The background state is based on the Jupiter 
model by \citet{French2012} and \citet{Nettelmann2012}, 
which uses {\it ab initio} simulations to
determine the equation of states for Hydrogen and Helium and 
to calculate the transport properties.
\Figref{fig:DensSigma} shows the normalized density profile $\rhob$ and 
electrical conductivity profile $\sigmab$ along with 
the profiles that have been assumed in the simulations. 
The transition to a metallic hydrogen state takes place 
at about $0.9\,r_J$. 
Jupiter's Steeply Decaying Conductivity Region (\SDCR) occupies 
the outer $10\,$\% in radius and is highlighted by a medium gray background color. 
Here, the conductivity decays increasingly rapidly with radius, 
eventually reaching super-exponential rates. 

The numerical models disregard the outer one percent in 
radius where the density gradient in steepest and thus 
very difficult to resolve numerically. 
The conductivity profile used in the simulations  
is a combination of a polynomial branch, 
\bel{eq:cond0}
\sigma = \sigma_i + 
(\sigma_m-\sigma_i) \left( \frac{r-r_i}{r_m-r_i}\right)^a 
\;\;\mbox{for}\;\; r<r_m
\ee
and an exponential branch,  
\bel{eq:cond1}
\sigma =  
\sigma_m \exp\left( \frac{r-r_m}{r_m-r_i}\frac{\sigma_m-\sigma_i}{\sigma_m}\right)
\;\;\mbox{for}\;\; r\ge r_m
\eep
Here $\sigma_i$ is the reference conductivity at the bottom
boundary 
$r_i$, $a$ is the exponential decay rate, and $r_m$ 
is the transition radius where both branches meet with 
$\sigma(r_m)=\sigma_m$. 
%The profiles in the dynamo simulations  
%G14, D18.20 and D18.55 use $r_r=r_i=0.2\,r_J$, $a=13$, $r_m=0.9$ and $%\sigma_m=\sigma_r/5$.
%The conductivity in D18.66 decays more rapidly with $a=16$. 

%LUCIA, PLOTTING SIGMA ACTUALLY SUGGESTS THAT $r_m=0.89$ AND A LARGE
% $r_m$ FOR THE NEW MODEL D18.66. IS THAT TRUE? PLEASE ALSO PROVIDE THE
% RAYLEIGH NUMBERS. I AM NOTE QUITE SURE ABOUT THE SCALING HERE.

While the {\it ab initio} simulations suggest that the conductivity already
slowly decreases in the metallic region, we assume a constant value here.
This keeps the magnetic Reynolds number at values that allow for dynamo action 
throughout this region \citep{Duarte2018}.
In the molecular region, our model profile decreases slower
than suggested by \citet{French2012} to ease the numerical calculations.
The total conductivity contrasts is about four orders of magnitude 
and we identify the \SDCR\ with the exponential branch \refp{eq:cond1}.

The mathematical formulation of the problem has been extensively
discussed elsewhere \citep{Jones2011,Wicht2018}. 
Detailed information can also be found in the online 
MagIC manual (http://magic-sph.github.io/). 
Here we only briefly introduce the essential equations. 

MagIC solves the non-dimensional Navier-Stokes equation, 
\begin{eqnarray} \label{eq:NS}
      \Ek\,\rhob\;\frac{d \Uv}{d t}
      & = &- \rhob\;\nabla\left(\frac{p}{\rhob}\right) + 2\rhob\;\Uv\times\uvz -
      \frac{\Ra^\prime\,\Ek}{\Pra\,\Di}\;\rhob\;\frac{\partial \Tb}{\partial r}\;S\,\uvr
\\ & & \nonumber
      + \frac{1}{\Pm_i}(\nabla\times\B)\times\B
      + \Ek\,\rhob\;\F_v \mcom 
\end{eqnarray}
the heat equation 
\bel{eq:Ener}
      \rhob\,\Tb\,\frac{d s}{d t}
          =  \frac{1}{\Pra}\nabla\cdot \left(\rhob\Tb\,\nabla s \right)
       + \frac{\Pra\,\Di}{\Ra^\prime} \left( q_\nu +  \frac{1}{\Pm_i^2\,\Ek} q_J \right)
       + q_s
\eec
the induction equation
\bel{eq:Dyn}
      \frac{\partial \Bv}{\partial t} = \nabla\times(\Uv\times\Bv )
      - \frac{1}{\Pm_i}\nabla\times(\lambdab\nabla\times\Bv ) 
\eec
the continuity equation 
\bel{eq:divU}
      \nabla\cdot (\rhob\Uv) = 0 
\eec
and the solonoidal condition for the magnetic field
\bel{eq:divB}
      \nabla\cdot\Bv= 0 
\eep

Here $p$ is a modified pressure that also includes 
centrifugal effects and accounts for disturbances in the 
gravity potential. Buoyancy variations are  
formulated in terms of the specific entropy $s$. Using the gradient of the 
normalized background temperature $\Tb$ in the buoyancy term
guarantees a consistent background gravity \citep{Wicht2018}. 
The three volumetric heating terms in \eqnref{eq:Ener} 
are viscous heating $q_\nu$, Joule or Ohmic heating $q_J$, and 
secular cooling $q_s$ \citep{Wicht2018}.

The dimensionless parameters ruling the system are 
the Ekman number 
\begin{equation}
      \Ek=\frac{\nu}{\Omega d^2} \mcom
\label{eq:ek}
\end{equation}
the modified Rayleigh number
\begin{equation}
      \Ra^\prime=\frac{\alpha_o \gb_o d^3 \Tb_o}{\nu\kappa_S } \frac{s_s}{c_p} = \Ra \frac{S_s}{c_p} \mcom
\label{eq:ra}
\end{equation}
the Prandtl number 
\begin{equation}
      \Pra=\frac{\nu}{\kappa_S} \mcom
\label{eq:pr}
\end{equation}
and the inner boundary magnetic Prandtl number 
\begin{equation}
      \Pm_i=\frac{\nu}{\lambda_i} \mdot
\label{eq:pm}
\end{equation}
 Here $\nu$ is the (homogeneous) kinematic viscosity,
$\Omega$ the rotation rate,   
$\alpha_o$ the outer boundary thermal expansivity , $\gb_o$ the outer-boundary gravity, $d=r_o-r_i$ the shell thickness, $\kappa_S$ the 
entropy diffusivity, $s_s$ the entropy scale, $c_p$ the 
heat capacity, 
$\lambda_i=1/(\mu_o\sigma_i)$ the inner boundary magnetic diffusivity,
$\uvr$ the radial unit vector, and $\uvz$ the unit vector in the 
direction of the rotation axis.   
The aspect ratio, another dimensionless parameter, has been fixed to $r_i/r_o=0.2$.

The modified Rayleigh number $\Ra^\prime$ is the product of the classical Rayleigh number $\Ra$ and the
dimensionless entropy scale $s_s/c_p$.
The dissipation number 
\bel{eq:di}
      \Di = \frac{d}{\Tb_o}\left(\frac{\partial \Tb}{\partial r}\right)_{r_o} = \frac{d \alpha_o \gb_o}{c_p}
\ee
is defined by the background temperature $\Tb$.

The equations have been non-dimensionalize by using the imposed entropy 
difference as entropy scale $s_s$, the difference between outer radius 
$r_o$ and inner radius $r_i$ as a length scale $d=r_o-r_i$, 
the viscous diffusion time $d^2/\nu$ as a time scale,
and $\sqrt{\Omega\mu_0\lambda_i\rhob_o}$ as a magnetic scale, 
where $\tilde{\rho}_o$ is the outer boundary reference density. 
We use entropy rather than temperate diffusion in the heat 
equation, which considerably simplifies the system \citep{Braginsky1995}.
In the above formulation, the dimensionless profiles $\rhob$, $\Tb$ and
$\lambdab$ carry the information on the
radial dependence of the background state.

\subsection{Poloidal/Toroidal Decomposition}

We use the common representation of the divergence free magnetic field 
by a poloidal and a toroidal contribution, 
\bel{eq:poltorB}
\Bv = \Bv^{(P)} + \Bv^{(T)} = \curl\curl\uvr\;\bpp + \curl\uvr\;\btp
\eec
where $b^{(P)}$ is the poloidal and $b^{(T)}$ the toroidal scalar potential.
The respective decomposition for the electric current density $\jv$ reads 
\bel{eq:poltorJ}
\jv = \jv^{(P)} + \jv^{(T)} = \curl\curl\uvr\;\jpp + \curl\uvr\;\jtp
\eec
where $\jpp$ denotes the poloidal and $\jtp$ the toroidal potential. 

Simple vector calculus yields 
\bel{eq:poltorJB}
\jv = - \nabla_H^2\,\jpp\;\uvr + 
\nabla_H \frac{\partial}{\partial r}\,\jpp + 
     \curl\,\uvr\;\jpp
\eec
where $\nabla_H$ is the horizontal component of the 
nabla operator.
The toroidal contribution obviously has no radial component. 
Radial derivatives appear only in the horizontal poloidal current density, 
which therefore dominates in the \SDCR, as we will show below. 

Ampere's law, $\curl \Bv=\mu \jv$, connects the  
poloidal (toroidal) magnetic field to the toroidal (poloidal) current density. 
Its radial component establishes a connection to the toroidal 
field potential: 
\bel{eq:Jr}
\mu j_r=- \mu \nabla_H^2\,\jpp\;\uvr = \uvr\cdot (\curl\Bv)= 
- \nabla_H^2 \btp\;\uvr
\eep 
The toroidal field at a given radius is thus determined 
by the radial currents flowing through the respective radial shell. 

\subsection{Selected Simulations}

We concentrate on closely analyzing two quite 
different dynamo simulations that both reproduce Jupiter's large scale field 
\citep{Duarte2018}. Model G14 has been introduced by
\citet{Gastine2014} while model D18 is listed as model number 20 
in \citet{Duarte2018}; \tabref{tab:parameters} compares 
their parameters.
Both dynamos share the background pressure, temperature, density, 
and electrical 
conductivity models (see \figref{fig:DensSigma}), 
use stress-free outer but rigid inner boundary condition, 
employ constant entropy boundary conditions, and are driven 
by heat coming in through the lower boundary. 
Model D18 uses a larger Ekman and larger magnetic Prandtl 
number. The Prandtl number is one in 
G14 but only $0.1$ in D18. A consequence is the  
the smaller scale field in D18 \citet{Duarte2018}.
 
\begin{figure}
\centering
\includegraphics[draft=false,width=0.80\textwidth]{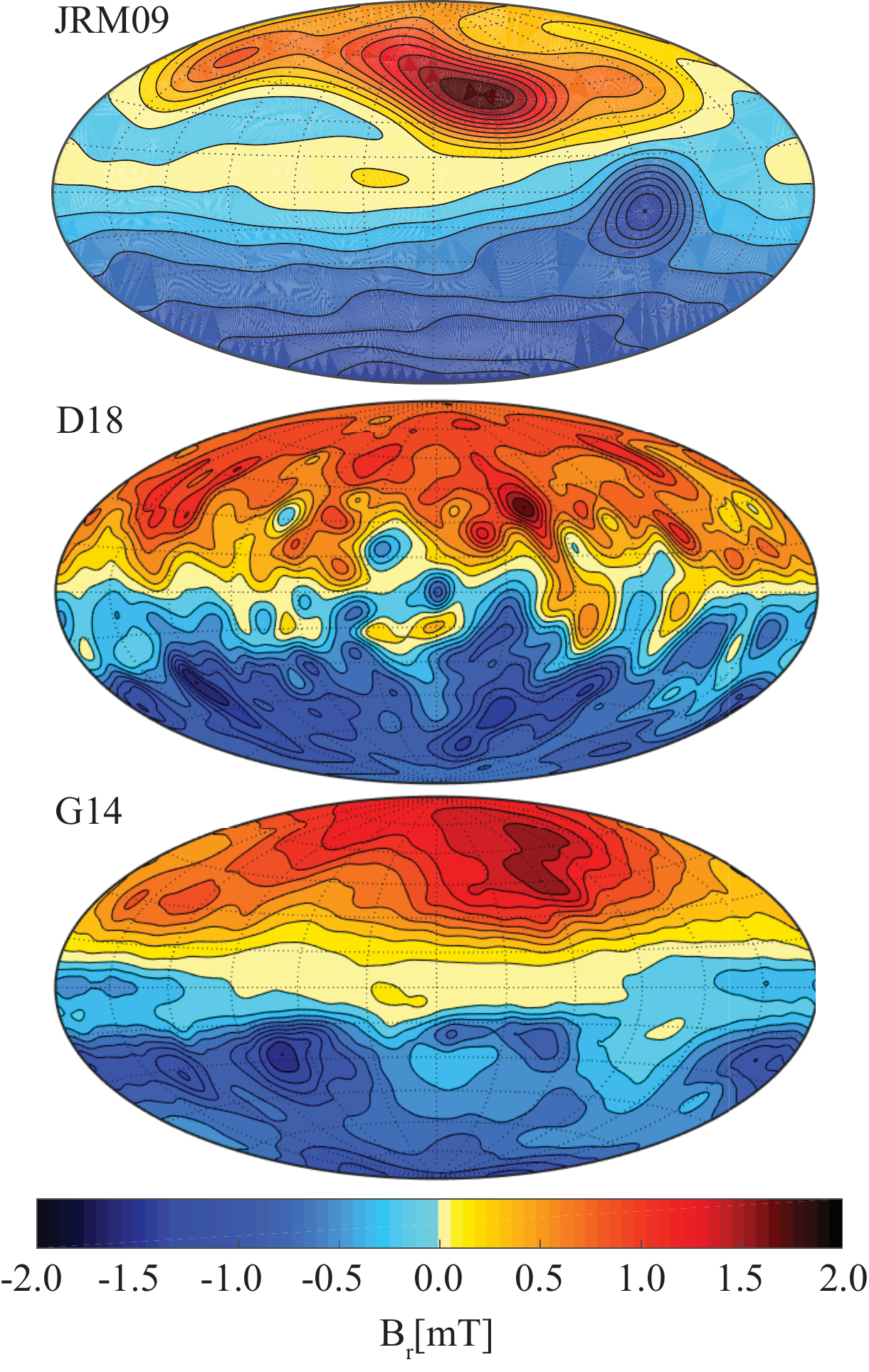}
\caption{Comparison of the surface radial magnetic field 
in the recent Jupiter field model JRM09 \citep{Connerney2018} with the 
field in two snapshots of the dynamo D18 and G14 closely 
explored here. G14 has been rescaled to $mT$ 
with the power based scaling \citep{Gastine2014,Duarte2018}.
Because this would yield a too strong field for D18, 
we have scaled the field strength of the snapshot to JRM09 
values for illustration purposes.  
}
\label{fig:Bcmb}
\end{figure}

\Figref{fig:Bcmb} compares 
the radial surface field for the two snapshots we will
continue to analyze throughout the paper with the recent 
Jupiter field model JRM09 by \citet{Connerney2018}. 
The non-dimensional fields in the computer models 
have been rescaled to reflect the Jovian parameters
using the methods explained in \citet{Gastine2014} 
and \citet{Duarte2018}. 

\begin{table}
\centering
\begin{tabular}{cccccccc}
Name & $\Ek$ & $\Ra$ & $\Pm$ & $\Pr$ & $\alpha$ & $r_m/r_o$ 
& $\sigma_m/\sigma_i$ \\
\hline
D18 & $10^{-4}$  & $6\tp{7}$ & $2$   & $0.1$& $13$ & $0.9$ & $0.5$ \\
G14 & $10^{-5}$  & $5\tp{9}$ & $0.6$ & $1$  & $13$ & $0.9$ & $0.5$ \\
\hline
\end{tabular} 
\label{tab:parameters}
\caption{Parameters of the numerical dynamo simulations explored here. 
Model G14 has first been discussed in \citet{Gastine2014} while 
D18 is model 20 from \citet{Duarte2018}. All simulations use the 
background density and temperature model following 
\citet{French2012}.}
\end{table}

\begin{figure}
\centering
\includegraphics[draft=false,width=0.7\textwidth]{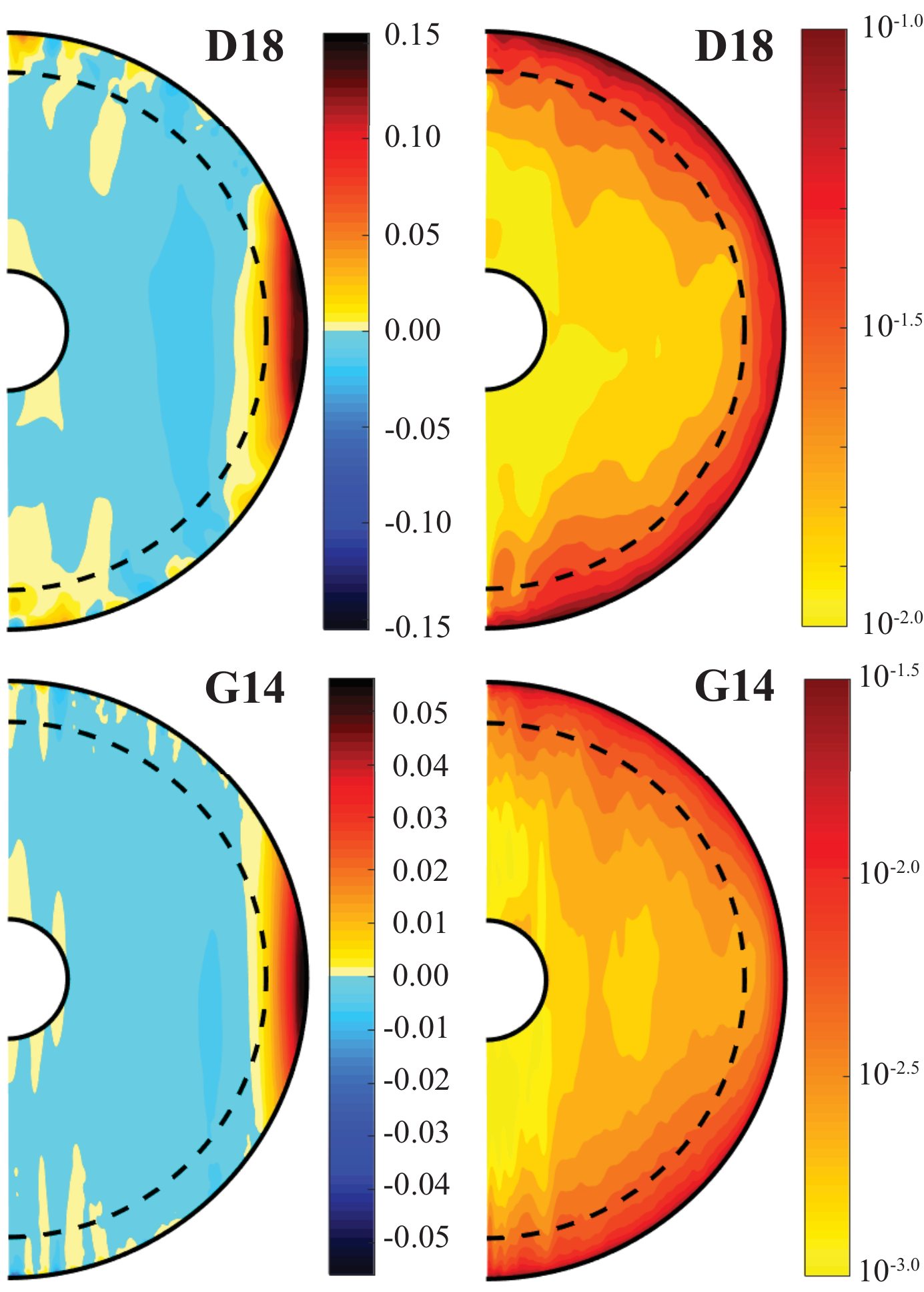}
\caption{Zonal flows ({\it left panels}) and 
rms non-axisymmetric flows ({\it right panels}) for 
dynamo D18 and dynamo G14 snapshots. Rossby number 
scales, $\Ro=\U / (d\Omega)$, have been used here.
}
\label{fig:ASFlow}
\end{figure}

\Figref{fig:ASFlow} illustrates the flow structure for 
both dynamos. They share the fact that the zonal flows 
({\it left panels}) 
show a pronounced equatorial jet but not the 
multitude mid to high latitude jets observed in 
Jupiter's or Saturn's cloud structure. These jets 
seem incompatible with dynamo generated  
dipole dominated magnetic fields in the numerical 
simulations \citep{Gastine2012a,Duarte2013}. 
The rms non-axisymmetric flow amplitude ({\it right panels} in \figref{fig:ASFlow}) increases with radius because 
of the decreasing density \citep{Gastine2012}.

\subsection{Magnetic Reynolds Numbers}

\begin{figure}
\centering
\includegraphics[draft=false,width=0.75\textwidth]{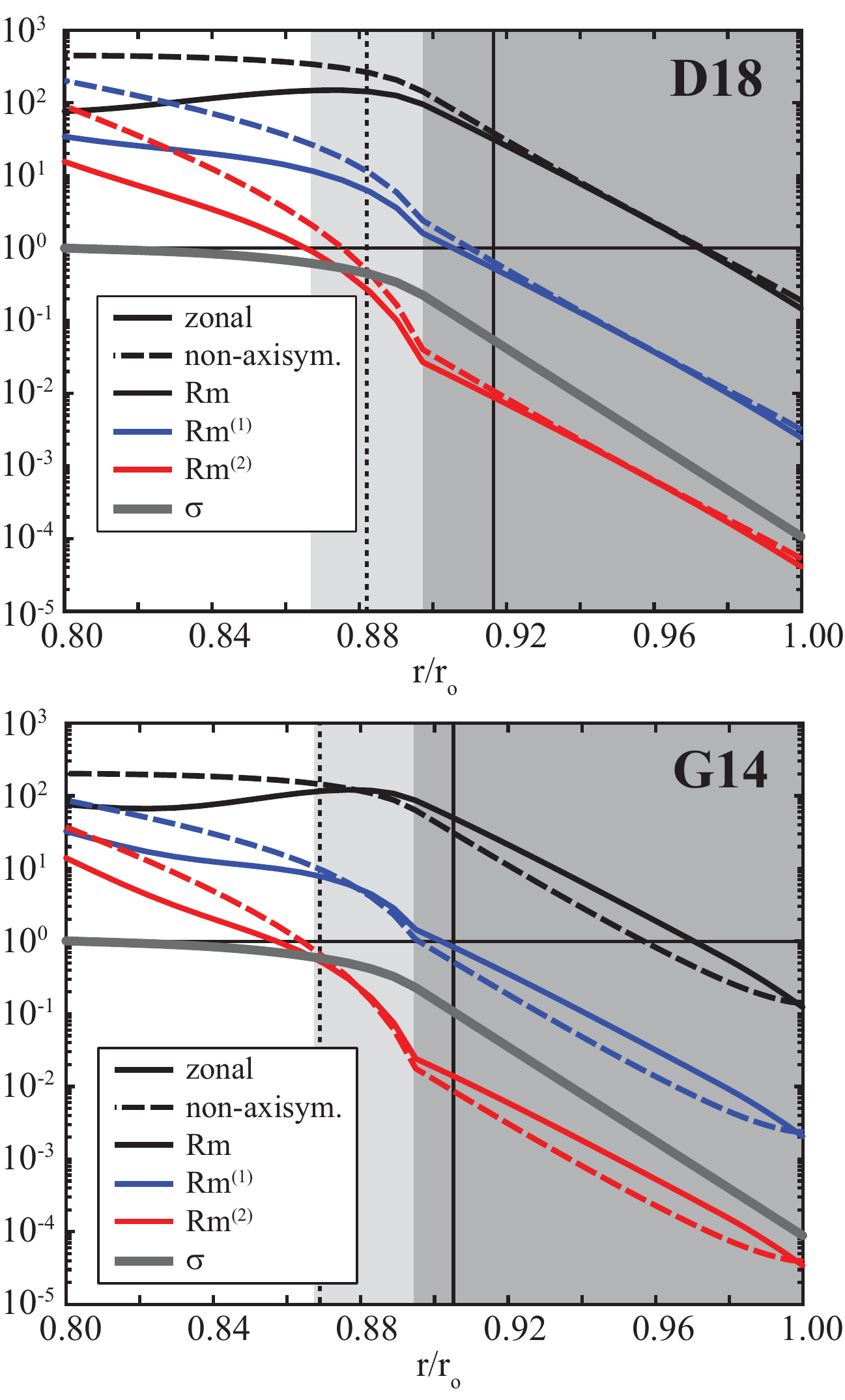}
\caption{Radial profiles of the three different magnetic Reynolds numbers for snapshots of dynamo models D18 and G14. {\it Solid (dashed) vertical lines} 
mark the radii where $\RmL$ ($\RmLS$) exceed unity. 
}
\label{fig:Rms}
\end{figure} 

As already discussed in the introduction, the strong 
decrease in the electrical conductivity in the \SDCR\  
requires modified magnetic Reynolds numbers. 
In addition to profiled obeying the classical definition,  
\bel{eq:Rm0}
\Rm(r) = \frac{\langle\Uv(r)\rangle\;d}{\lambda(r)}
\eec
we will rely on  
\bel{eq:Rm1}
\RmL(r)= \frac{\langle\Uv(r)\rangle\;d_\lambda}{\lambda(r)}
\eep
and also 
\bel{eq:Rm2}
\RmLS(r) = \frac{\langle\Uv(r)\rangle\;d_\lambda^2}{\lambda(r)\;d}
\eep
Here, $d_\lambda = \lambda /(\partial \lambda / \partial r)$ 
is the magnetic diffusivity scale height, and 
$\langle\U(r)\rangle$ refers to the rms velocity 
on the sphere of radius $r$. 
In general, angular brackets indicate the spherical rms 
\bel{eq:Jrms}
\langle f(r) \rangle \:= \frac{1}{4\pi}\;
\left(\int_{-1}^{1} d x \int_{0}^{2\pi} 
d \phi f^2 \right)^{1/2}
\ee
throughout the paper, 
where $x=cos(\theta)$, $\theta$ is the colatitude 
and $\phi$ the longitude. 

\Figref{fig:Rms} compares the different magnetic 
Reynolds number profiles for the D18 and G14 snapshots, 
separating between zonal flow and non-axisymmetric flow 
contributions. Axisymmetric latitudinal and radial flows 
are typically much weaker and have thus been neglected here.  
While zonal and non-axisymmetric flows
have comparable amplitudes  in the \SDCR\ of dynamo D18, 
zonal flows are somewhat more pronounced in model G14 
because of the smaller Ekman number. 
In the exponential branch of the conductivity profiles,  
the diffusivity 
scale height assumes a constant value $d_\lambda=0.168$.
This explains why the different \Rm\ profiles in
\figref{fig:Rms} are parallel in the \SDCR. 

%Mean field dynamo theory shows that toroidal field 
%is either created by shear in the zonal flow $\Uz$ or 
%helical flow structures on the non-axisymmetric flow $\UNA$. 
%These related processes are called $\Omega-$effect and
%$\alpha$-effect, respectively, and we accordingly distinguish 
%magnetic Reynolds number based on an rms 
%$\Rm_Z$ and $\Rm_{NA}$ by using the 
%the respective rms velocities. 
%Since the $\alpha$ process is the only possibility 
%to produce poloidal from toroidal field, only the 
%latter magnetic Reynolds number is of interest 
%when considering mean poloidal field generation. 

\section{Dynamo Action in the \SDCR}
\label{sec:Results}

Ohm's law, 
\bel{eq:Ohm}
\jv = \sigma\left(\Uv\times\Bv + \Ev\right)
\eec
states that the electric currents are driven by 
induction due to flow acting 
on magnetic field or by the electric field $\Ev$.
Since both effects scale with $\sigma$, the current density $\jv$ 
has to decays in the \SDCR. 

\begin{figure}
\centering
\includegraphics[draft=false,width=0.75\textwidth]{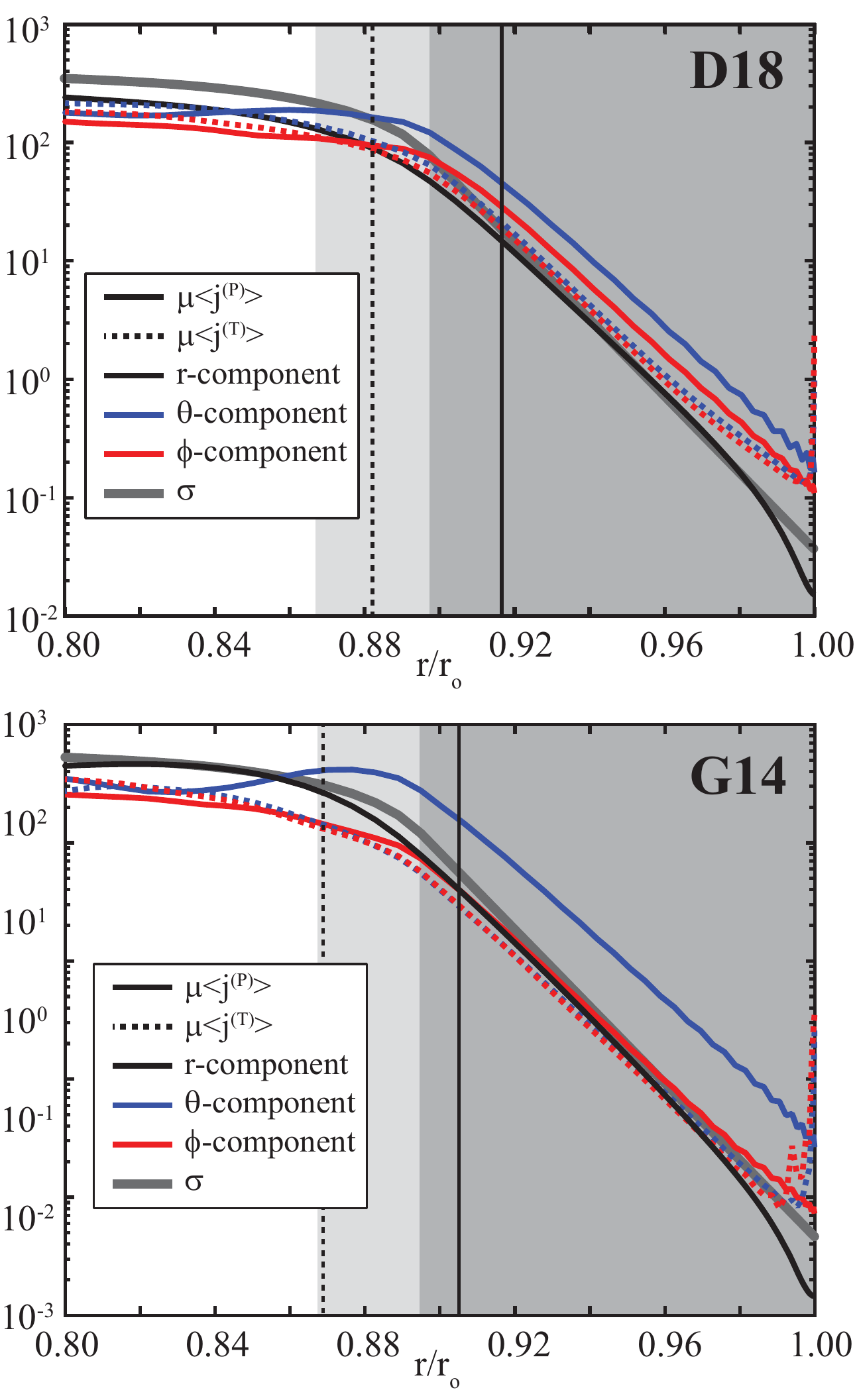}
\caption{Radial profiles for the rms values of different current 
contributions. A scaled electrical conductivity profile is 
shown in gray. Background colors indicate the region where 
the conductivity is smaller than $50$\% of its reference 
values at $r_i$ ({\it light gray background}) and the \SDCR\ where is decays 
exponentially ({\it mid gray background}). Top and bottom panels show 
snapshots for dynamos D18 and G14, respectively. 
{\it Solid (dashed) vertical lines} 
mark the radii where $\RmL$ ($\RmLS$) exceed unity.
}
\label{fig:Jcomp}
\end{figure} 

\Figref{fig:Jcomp} illustrates the decay of the different current density components for the selected snapshots in dynamos D18 ({\it top panel}) 
and G14 ({\it bottom panel}). 
Shown are radial profiles of the rms values over spherical surfaces,  
indicated by angular brackets.
The poloidal current ({\it solid lines}), more specifically its latitudinal component ({\it solid blue}), dominates in the \SDCR. 
The poloidal current is at least two times larger 
than the toroidal current in model D18. The difference is even more 
pronounced in model G14, where the poloidal current is five times 
stronger than its toroidal counterpart, likely due to the stronger 
zonal flows.
Radial currents ({\it solid black}) are effectively blocked off by the 
conductivity gradient and are thus generally small. 

As expected, the decrease of the rms current density 
in the \SDCR\ is mostly dictated by the electrical
conductivity profile ({\it thick gray line in \figref{fig:Jcomp}}) 
with local induction effects providing some moderation. 
Below the \SDCR, the current increases much more smoothly with depth and 
never exceeds twice the value reached at the 
bottom of the \SDCR. The simple 
conductivity ruled gradients changes at a depth somewhere 
between $\RmL=1$ and $\RmLS=1$ where more classical dynamo
action kicks in.

In the simulations, we assume that the region $r>r_o$ is 
electrically insulating. The conditions that are 
gradually approached with increasing radius in the \SDCR\ 
are finally abruptly enforced at $r_o$. \Figref{fig:Jcomp} illustrates 
that this leads to a thin magnetic boundary layer where the simple 
dependence on $\sigma$ also brakes down. Another problem is that 
the very small currents in this very outer region cannot be 
calculated precisely enough, since we had to calculated from 
single precision magnetic field values stored in the snapshots. 

In order to quantify the locally induced current-related poloidal field, 
we downward continue $\tilde{\Bv}$ below $r_o$ 
according to the characteristic radial dependence of 
a potential field: 
\bel{eq:Potr}
 \tilde{\Bv}_\ell(r) = 
 \left( \frac{r_o}{r} \right)^{\ell+2}\;\Bv_\ell(r_o)
\eep
In the \SDCR\ where local currents are small, 
the $\Bv$ remains close to $\tilde{\Bv}$. 
The difference 
\bel{eq:Bprime}
    \hat{\Bv}^{(P)}(r)=\Bv^{(P)}(r)-\tilde{\Bv}(r) 
\ee
provides a measure for the poloidal field produced by the local 
currents flowing beyond radius $r$. For simplicity, we will refer 
to $\hat{\Bv}$ as the non-potential poloidal field in the following. 

\begin{figure}
\centering
\includegraphics[draft=false,width=0.75\textwidth]{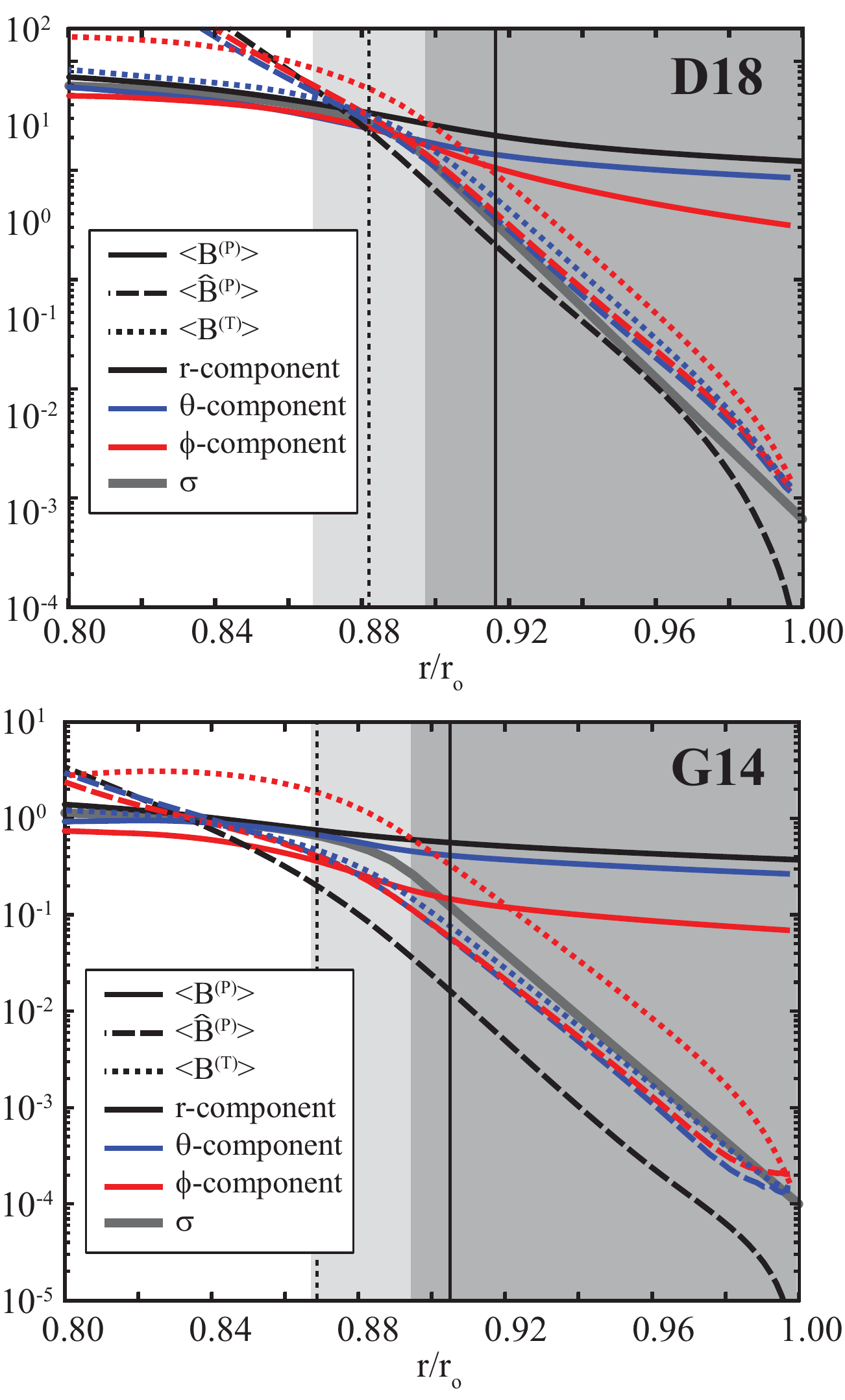}
\caption{Rms values of different magnetic field contributions. 
See caption of \figref{fig:Jcomp} for more explanation.
{\it Solid (dashed) vertical lines} 
mark the radii where $\RmL$ ($\RmLS$) exceed unity. 
Dimensionless quantities
shown.
}
\label{fig:Bcomp}
\end{figure} 

\Figref{fig:Bcomp} illustrates the radial dependence 
of different rms magnetic field contributions. 
The toroidal magnetic field connected to the poloidal currents 
dominates the non-potential field in the \SDCR. 
The shear due to the zonal flow ($\omega$-effect) produces 
a particularly strong azimuthal toroidal field, 
an effect that is more pronounced in dynamo G14 than in D18. 
Since the rms azimuthal potential field is 
particularly weak, the locally induced field reaches a an amplitude 
comparable to $\langle\tilde{\B}_\phi\rangle$  already within the \SDCR. 
Using the downward continuation \refp{eq:Potr} to deduce 
$\hat{\Bv}$ stops to make sense when $\tilde{\Bv}$ becomes of the same 
order as $\Bv$. \Figref{fig:Bcomp} shows that $\hat{\Bv}$ 
can even exceeds $\Bv$ below the \SDCR. 
As we will further discuss below, the radius where 
$\RmL$ exceeds unity, marked by a solid vertical line 
in \figref{fig:Bcomp}, is the depth where out approach brakes 
down. 

%The boundary condition applied at $r_o$ 
%leads to a faster decay in the radial current but also in the 
%toroidal field and the non-potential poloidal field 
%in the very outer two percent in radius. 
%This region should thus not be interpreted for 
%determining typical magnetic effects in the \SDCR. 

\begin{figure}
\centering
\includegraphics[draft=false,width=0.75\textwidth]{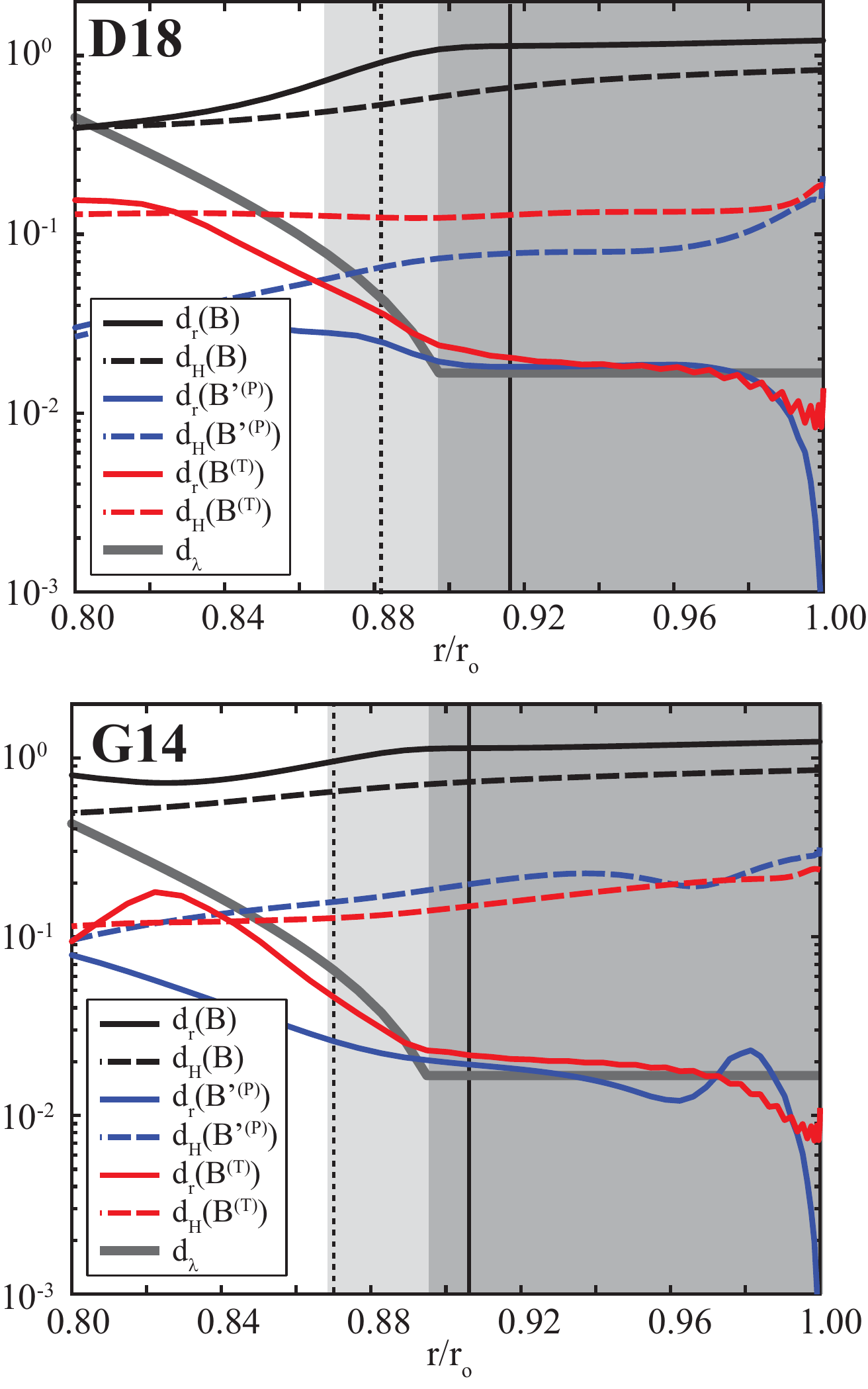}
\caption{Radial ({\it solid}) and horizontal ({\it dashed}) 
rms scales of different field contributions. 
See caption of \figref{fig:Jcomp} and the main text for more explanation.
}
\label{fig:Bscales}
\end{figure} 

Radial and horizontal magnetic field length scales, based on the respective 
derivatives of the field potentials, are illustrated in \figref{fig:Bscales}.
For example, the radial scale of the poloidal field is estimated via 
\bel{eq:scale}
d_r (B^{(p)}) = \frac{\langle b^{(P)} \rangle}
                     {\langle \partial \bpp / \partial r \rangle}
\eep
Horizontal scales have been calculated using the 
square root of $\langle\nabla_H^2 g\rangle$. 
Between $r=0.90\,r_o$ and $r=0.97\,r_o$, the radial scale of 
$\hat{Bv}$ and the toroidal field is indeed very similar to the diffusive 
scale height $d_\lambda$, which confirms that the conductivity 
profile determines the non-potential magnetic field gradient. 
The horizontal length scales of both non-potential field contributions are 
significantly smaller than the respective scales of the 
total field (or the potential field) but still an order of magnitude 
larger than the radial scales. 

The separate evolution equations for the toroidal and poloidal field 
potentials solved by MagIC are the radial component of the 
dynamo equation and the radial component of its curl 
\citep{Christensen2015}, 
\bel{eq:dtpol}
\nabla_H^2\;\dot{b}^{(P)} =
- \uvr\cdot\left[ \curl \left( \Uv\times\Bv \right) \right] + 
\lambda\;\nabla_H^2\,
\left[ \left( \frac{\partial}{\partial r} \right)^2 + 
\nabla_H^2 \right]\;\bpp
\ee
and
\bel{eq:dttor}
\nabla_H^2\;\dot{b}^{(T)}=
- \uvr\cdot\left[\curl\curl\left(\Uv\times\Bv\right)\right] + 
\lambda\;\nabla_H^2\,
\left[\;\left(\frac{\partial}{\partial r}\right)^2 + \nabla_H^2 
\;+\;\frac{1}{d_\lambda}\,\frac{\partial}{\partial r}\right]\;\btp
\eep
Only the toroidal field 
equation directly includes the magnetic diffusivity scale height.

%ONLY FOR FIGURES:
%\bel{eq:dt}
%\nabla_H^2\;\dot{b}^{(P)} =
%- \uvr\cdot\left[ \curl \left( \Uv\times\hat{\Bv}^{(P)} \right) \right] + 
%\lambda\;\nabla_H^2\,
%\left[ \left( \frac{\partial}{\partial r} \right)^2 + 
%\nabla_H^2 \right]\;\bpp
%\ee
%\be
%\uvr\cdot\left[ \curl \left( \Uv\times\Bv^{(T)} \right) \right] 
%\ee
%\be
%\uvr\cdot\left[ \curl \left( \overline{\U}_\phi\times\Bv\right) \right] 
%\ee
%\be
%\lambda\;\nabla_H^2\;\left( \frac{\partial}{\partial r} \right)^2\;\bpp
%\ee
%\be
%\lambda\;\nabla_H^2\;\nabla_H^2\;\bpp
%\ee

%\bel{eq:dtt}
%\nabla_H^2\;\dot{h} =
%- \uvr\cdot\left[\curl\curl\left(\overline{\U}_\phi\times\Bv\right)\right] + 
%\lambda\;\nabla_H^2\,
%\left[\;\left(\frac{\partial}{\partial r}\right)^2 + \nabla_H^2 
%\;+\;\frac{1}{d_\lambda}\,\frac{\partial}{\partial r}\right]\;\btp
%\eep
%\be
%\lambda\;\nabla_H^2\,\left(\frac{\partial}{\partial r}\right)^2\;\btp
%\ee
%\be
%\lambda\;\nabla_H^2\,\nabla_H^2\;\btp
%\ee
%\be
%\lambda\;\nabla_H^2\,\frac{1}{d_\lambda}\,\frac{\partial}{\partial r}\;\btp
%\ee

\begin{figure}
\centering
\includegraphics[draft=false,width=0.80\textwidth]{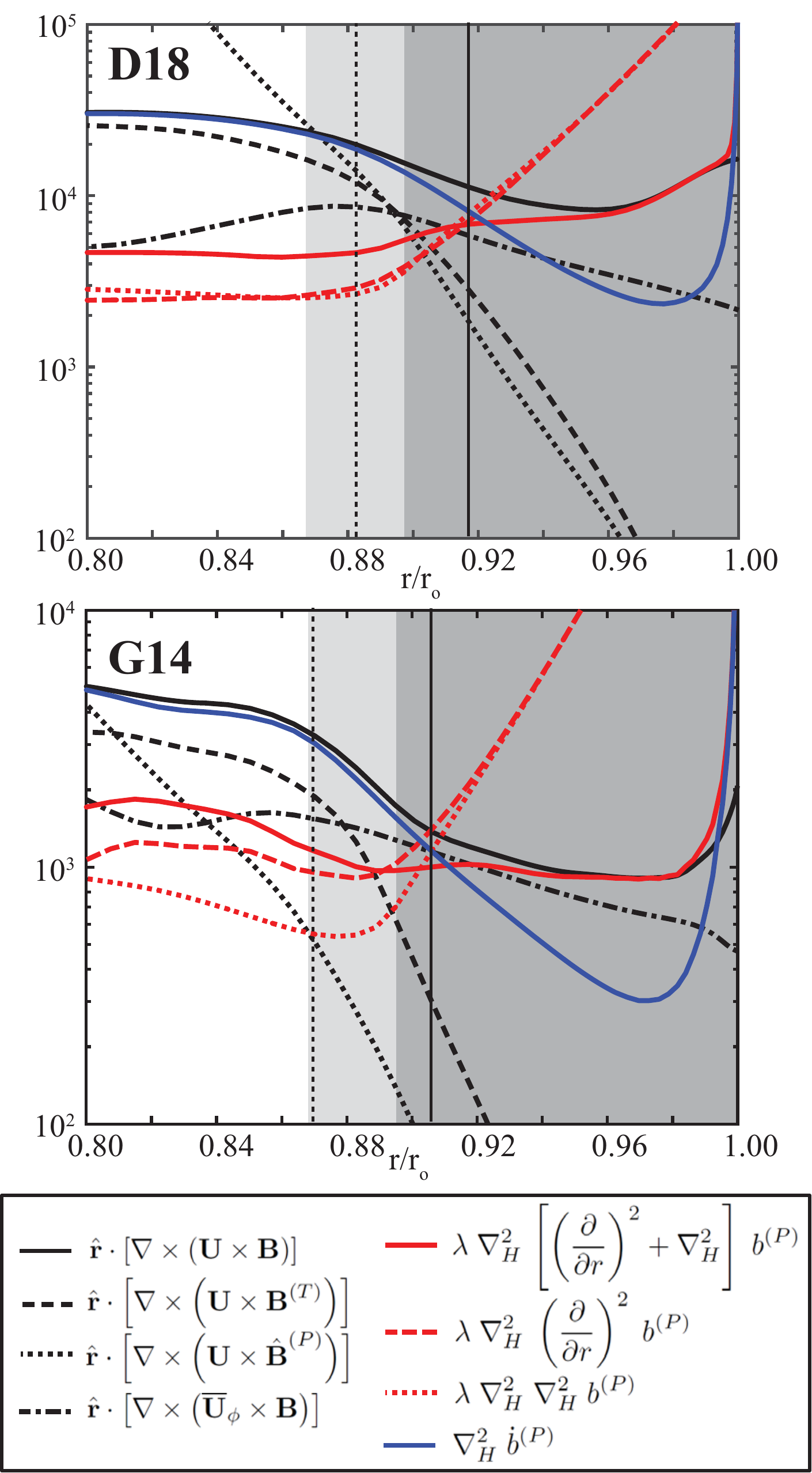}
\caption{Horizontally averaged rms values of the different contributions 
to the poloidal dynamo equation for the snapshots in dynamos 
D18 ({\it top}) and G14 ({\it bottom}).
The vertical lines mark where $\RmL$ ({\it solid}) and 
$\RmLS$ ({\it dotted}) reach unity. Dimensionless quantities
shown.}
\label{fig:dBpol}
\end{figure}

\Figref{fig:dBpol} shows radial profiles of the rms values
of the different contributions in the poloidal evolution equation 
for the D18 and G14 snapshots. 
Very close to the outer boundary beyond $r\approx 0.98\,r_o$, 
the use of single precision snap-shot data limits the 
quality of the second radial derivative required for the 
diffusive contributions. The balance should thus 
not be interpreted in this region. 

We have separated the diffusive term 
into two contributions involving 
radial ({\it red dashed line}) and only horizontal ({\it red dotted}) derivatives. 
Both increase with $\lambda$ towards the outer boundary but 
also progressively cancel each other since the 
field approaches a potential field. 
What remains is the diffusive term for the 
non-potential field ({\it red solid}), which is 
mostly balanced by the induction term ({\it black solid}). 
Consequently, magnetic field variations ({\it blue}) 
become comparatively small in the highly diffusive \SDCR.

\Figref{fig:dBpol} also illustrates that induction due to 
zonal flows ({\it black dashed-dotted})
%\bel{eq:uzi}
%\uvr\cdot\left[ \curl \left( \overline{\Uv}_\phi\times\Bv \right) \right]
%\eec
is sizable for the poloidal field evolution. This 
describes pure advection of the background field $\tilde{\Bv}$. 
Induction due to flow acting on the toroidal field 
({\it black dashed})
%\bel{eq:ubt}
%\uvr\cdot\left[ \curl \left( \Uv\times\Bv^{(T)} \right) \right]
%\eec 
or the non-potential poloidal field ({\it black dotted}) 
%\bel{eq:ubZ}
%\uvr\cdot\left[ \curl \left( \Uv\times\hat{\Bv}^{(P)} \right) \right]
%\eec
on the other hand, remain minor contributions in the \SDCR.  

\begin{figure}
\centering
\includegraphics[draft=false,width=0.80\textwidth]{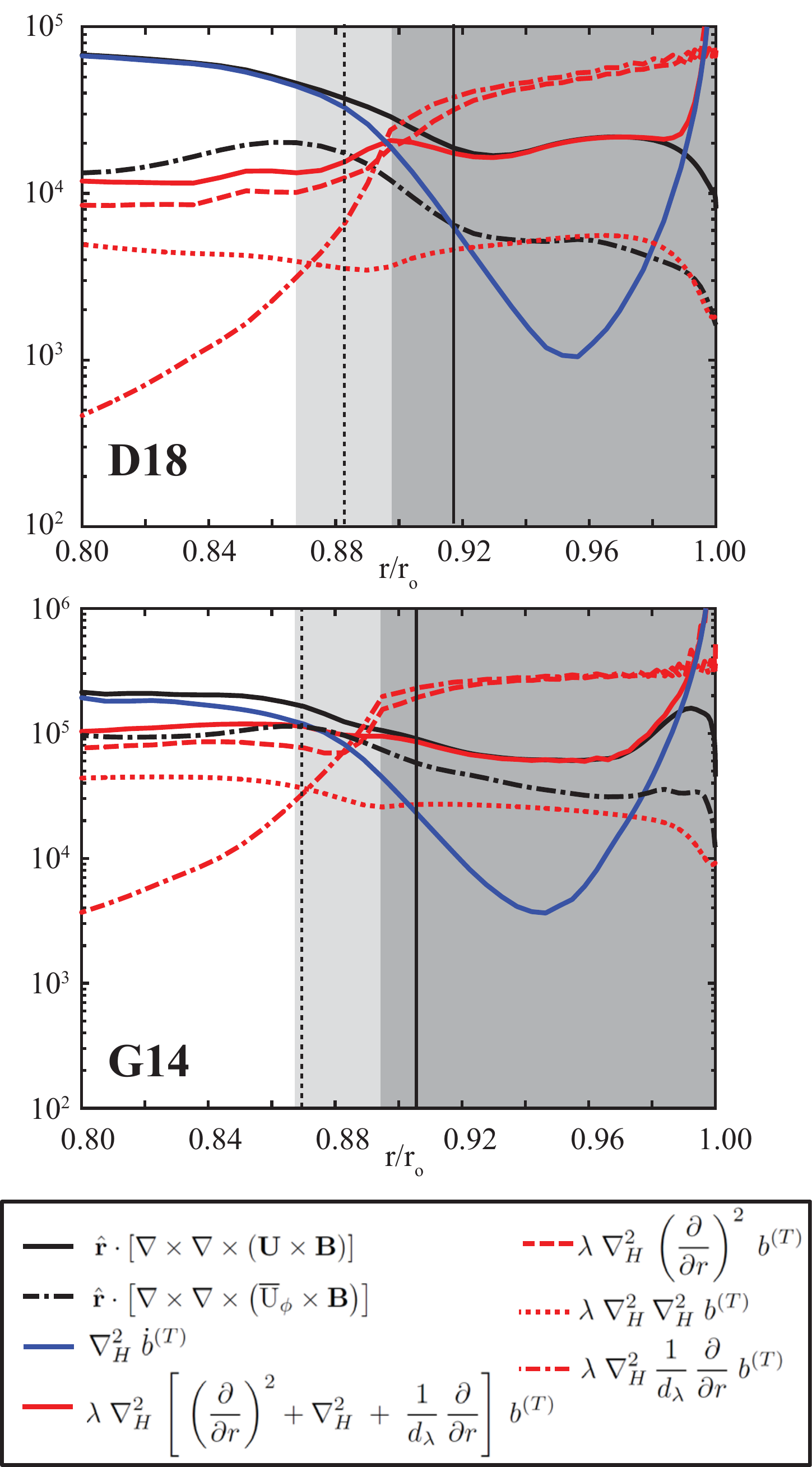}
\caption{Horizontally averaged rms values of the different contributions 
to the toroidal dynamo equation for the snapshots in dynamos 
D18 ({\it top}) and G14 ({\it bottom}).
The vertical lines mark where $\RmL$ ({\it solid}) and 
$\RmLS$ ({\it dotted}) reach unity. Dimensionless quantities
shown.}
\label{fig:dBtor}
\end{figure}

For the rms balance in the toroidal field evolution, 
illustrated in \figref{fig:dBtor}, 
time variations are even less important than for the poloidal counterpart. 
Note that the two dominant diffusive terms that involve radial 
derivatives ({\it red dashed and dash-dotted lines}) 
cancel to a high degree. This is simply a consequence 
of the fact that the toroidal field vanishes with $\sigma$ so that 
\bel{eq:diffT}
  \frac{\partial}{\partial r}\;\btp \approx - \frac{\btp}{d_\lambda}
\eep
Zonal flows clearly dominate the toroidal field induction, 
in particular for dynamo G14. 

Vertical lines in \figref{fig:dBpol} and \figref{fig:dBtor} 
again mark where the magnetic Reynolds numbers $\RmL$ and 
$\RmLS$ reach unity. The assumption of quasi stationary 
seems to be limited to the region $\RmL>1$. 
Where $\RmLS$ reaches unity at a somewhat greater depth, 
the induction term already clearly dominates diffusion 
and the dynamics is decisively non-stationary. 
\section{Estimating the Dynamo Action}
\label{sec:Estimates}

The analysis of dynamo simulations has shown that the dynamo
action in the \SDCR\  is dominated by diffusive effects 
and thus obeys a quasi-stationary dynamics. 
In this section, we exploit this simplicity and derive  
estimates which could be used to estimate the dynamo 
action in the \SDCR\ of planets. 
%Besides a surface field model, these estimates require a conductivity 
%profile and a flows model. 

\subsection{Estimating the Electric Currents}
The decent balance between diffusion and induction in the 
dynamo equation suggests   
\bel{eq:dBH}
\left( \curl\sigma \jv \right) \approx - \curl \left(\Uv\times\tilde{\Bv} \right)
\eep
When using the fact that radial derivatives dominate in the \SDCR, 
this leads to the radial integral estimate $\Iv_H$ for the horizontal
currents used by \citep{Liu2008}: 
\bel{eq:UxBI}
\jv_H(r) \approx \Iv_H(r) = \sigma\;\int_{r}^{r_o}\,d r^\prime\;
\uvr\times\left[ \curl \left( \Uv\times\tilde{\Bv} \right) \right]_H\;+\;C_H
\eep
Here $C_h$ stands for the integration constants which 
guarantees that the current vanishes for $r=r_o$. 

A simpler estimate can be based on Ohm's law. Balance \refp{eq:dBH} 
already suggest that electric field contributions associated to variation
in the magnetic field via the Maxwell-Faraday law, 
$\partial \Bv/ \partial t = - \curl\Ev$, are small. A second possible contribution
are potential field gradient, $\E=-\nabla V$. 
At least the latitudinal component of $\Uv\times\Bv$ clearly dominates the
respective electric field contribution and the 
fast zonal flows likely also play a role here. This suggest to use the 
simplified Ohm's law for a fast moving conductor:
\bel{eq:UxB}
\jv \approx \sigma\;\Uv\times\tilde{\Bv}
\eep

\begin{figure}
\centering
\includegraphics[draft=false,width=0.75\textwidth]{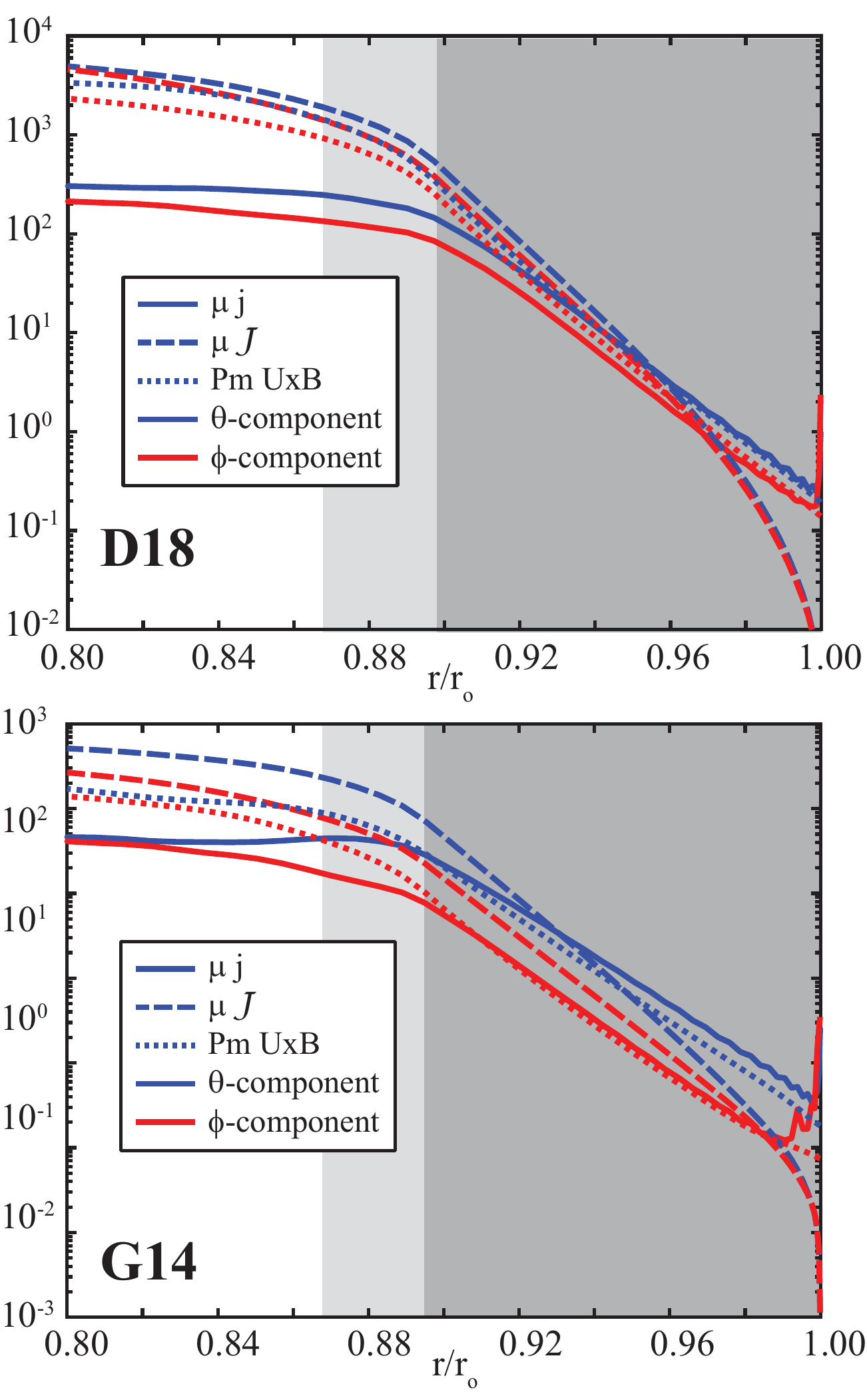}
\caption{Radial profiles of the rms horizontal current
densities along with two different estimates. Dimensionless quantities
shown.}
\label{fig:Jcompest}
\end{figure} 

\Figref{fig:Jcompest} illustrates that the rms horizontal 
currents in the \SDCR\ are indeed close to the rms value of 
$\sigma \Uv\times\Bv$ ({\it dotted lines}) with 
deviations below $50$\% for $r>0.90\,r_o$. 
The integral estimate $\Iv_H$, shown as 
dashed lines in \figref{fig:Jcompest}, provides a somewhat 
inferior estimate but nevertheless correctly 
captures the order of magnitude. 
Below the \SDCR, however, $\sigma \Uv\times\Bv$ is much 
larger than $\jv$ and the electric field 
(or magnetic field variations) can definitely not be neglected.

While \eqnref{eq:UxB} convincingly approximates the horizontal 
currents densities, it severely overestimates the much weaker 
radial component. 
%In particular the 
%zonal flows contributions to the radial component of 
%$\Uv\times\Bv$ are effectively balanced by an electric field for this 
%weaker current contribution. 
We can, however, use the fact that the currents are predominantly 
poloidal.
% so that 
%\bel{eq:JPH}
%\jv \approx \nabla_H \frac{\partial}{\partial r} i \approx \sigma\;
%\left(\Uv\times\Bv\right)_H
%\eec
%where the index $H$ denotes the horizontal components. 
Taking the horizontal divergence of \eqnref{eq:UxB} yields
\bel{eq:DJPH}
\nabla_H^2 \frac{\partial}{\partial r} j^{(p)} \approx \sigma\;
\nabla_H\cdot\left(\Uv\times\Bv\right)_H
\eep
This suggest the alternative integral estimate 
\bel{eq:JIR}
J_r (r) \approx \I_r(r) = -\int_{r}^{r_o} d r^\prime\; 
\sigma\;\nabla_H\cdot\left(\Uv\times\Bv\right)_H + C_r
\eec
where $C_r$ is an integration constant used to  
assure that $J_r$ vanishes at $r_o$. 
Since the gradient in $\sigma$ will dominate the integral, 
the simplified expression 
\bel{eq:JIRE}
J_r \approx d_\lambda 
\sigma\;\nabla_H\cdot\left(\Uv\times\tilde{\Bv}\right)_H 
\ee
provides a very decent estimate.

For estimating the rms electrical current, we rely on the 
estimate for the dominant horizontal component.
Using background field $\tilde{\Bv}$ and ignoring the cross product in \eqnref{eq:UxB} leads to    
\bel{eq:JrmsS}
\langle\Jv\rangle \approx \sigma\;\langle\Uv\rangle\;
\langle\tilde{\Bv}\rangle = \frac{\Rm}{\mu\,d}\;\langle\tilde{\Bv}\rangle
\eec
while only slightly degrading the estimate. 
To quantify its quality, we calculate the ratio 
of the estimate to the true value,
\bel{eq:Jratio}
R_{J} = \frac{\sigma\,\langle\Uv\rangle\,\langle\tilde{\Bv}\rangle}
                             {\langle\Jv\rangle}
\eep
%This is compared with the respective ratio for the 
%integral estimate $\Iv_H$ is \figref{fig:Jratio}.  

For the radial component, the ratio based on 
\eqnref{eq:JIRE} reads
\bel{eq:JrRatio}
R_{Jr} = d_\lambda\,\sigma
\frac{\langle\nabla_H\cdot\left(\Uv\times\tilde{\Bv}\right)_H\rangle}
{\langle\J_r\rangle}
\eep
Ignoring cross product and divergence in \eqnref{eq:JIRE} 
yields the simpler estimate 
\bel{eq:JIRES}
\langle J_r \rangle  \approx \sigma\;
\frac{d_\lambda}{d} \langle\Uv\rangle \langle\tilde{\Bv}\rangle 
= \frac{\RmL}{\mu\,d}\;\langle\tilde{\Bv}\rangle
\ee
with the respective ratio
\bel{eq:JrsRatio}
\hat{R}_{Jr} = \frac{\sigma d_\lambda\,\langle\Uv\rangle\,\langle\tilde{\Bv}\rangle}{d\,\langle\Jv\rangle}
\eep

\begin{figure}
\centering
\includegraphics[draft=false,width=0.75\textwidth]{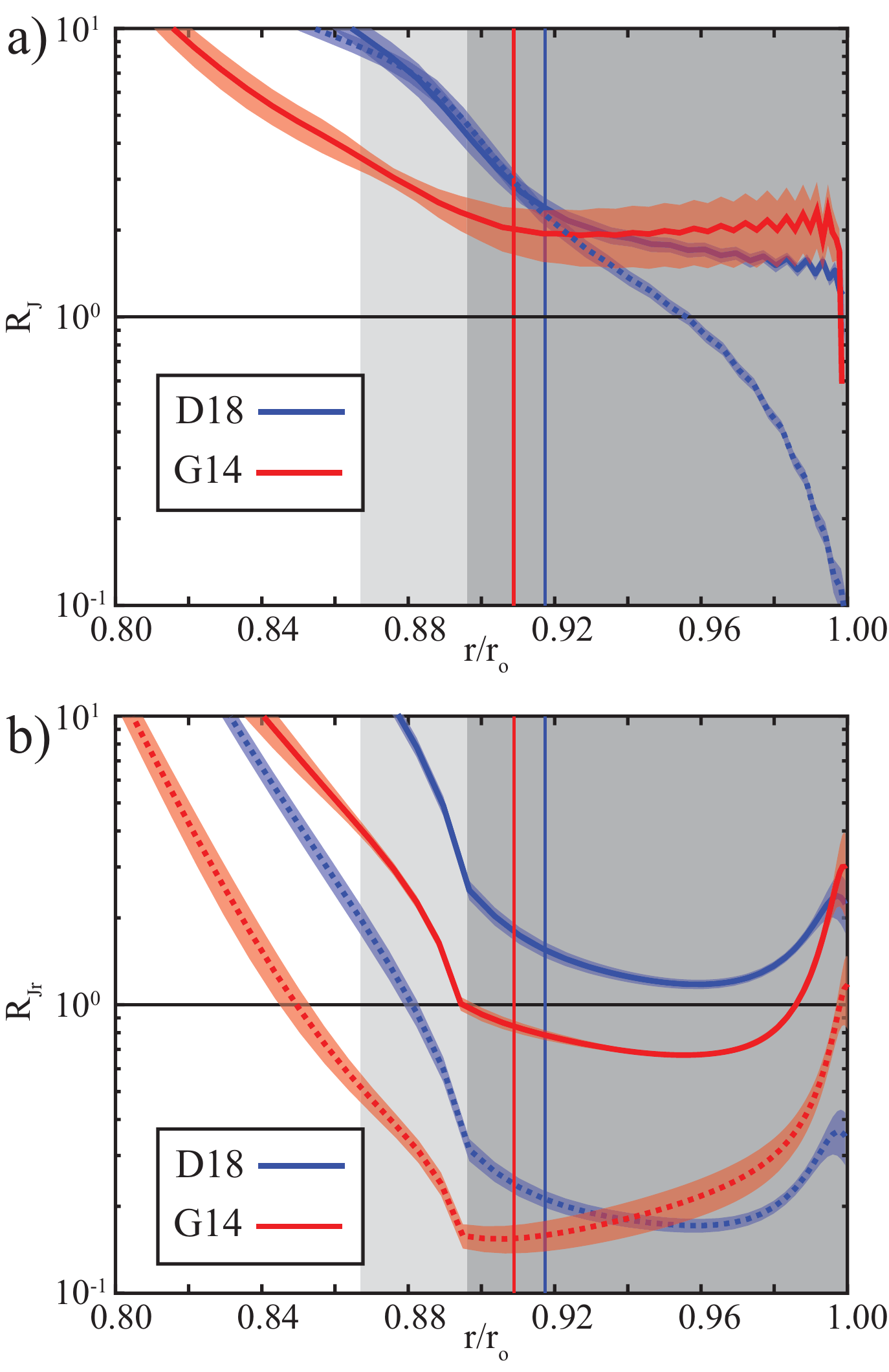}
\caption{Ratio of estimates and true rms electric current density.
The results have been averaged over 14 snapshots for dynamo D18 
and 10 snapshots for dynamo G14. 
Transparent background stripes with a width of two standard deviations 
illustrate the variability. The vertical lines mark where 
$\RmL$ ({\it solid}) or $\RmLS$ ({\it dotted}) exceed one. {\it Panel a)} shows ratio 
\refp{eq:Jratio} for the simplified estimate based on Ohm's 
law ({\it solid lines}) and an respective ratio 
for the integral estimate \refp{eq:UxBI} $\Iv_H$ ({\it dotted lines}). 
{\it Panel b)} compares ratios \refp{eq:JIRE} ({\it full lines}) and \refp{eq:JIRES} {(\it dotted lines)}
for the radial current contributions.}
\label{fig:Jratio}
\end{figure}

\Figref{fig:Jratio} shows the mean values and standard 
deviations for the different ratios when using 
14 snapshots for D18 and $10$ for G14. 
When using Ohm's law for a fast moving conductor, the rms currents 
are overestimated by factors 
between $1.5$ and $2.5$ for $r>0.92 r_o$ 
({\it solid lines} in \figref{fig:Jratio}a). The quality of the 
integral estimate, however, varies significantly with depth and 
provides reasonable values for $0.91\,r_o < r < 0.96\,r_o$. 
Estimates for the smaller radial current are of generally 
high quality when including the divergence 
in \refp{eq:JIRE} ({\it solid lines in \figref{fig:Jratio}b}). 
Ignoring the divergence, however, leads to values that are about 
an order of magnitude too low 
({\it dotted lines in \figref{fig:Jratio}b}). 
 
\Figref{fig:Jratio} also suggest that the estimates 
remain more or less valid throughout the whole \SDCR.  
The assumption of a dominant dominant background potential
field but also for the quasi-stationary dynamo action 
roughly holds for $\RmL\ge 1$ and the radii where $\RmL$ reaches unity 
have been marked 
with vertical lines in \figref{fig:Jratio}. 
The validity of our estimates extends to slightly greater 
depth, reaching $\RmL\approx 2$ 
and $\RmL\approx 1.5$ at the bottom of the \SDCR\ for 
dynamos D18 and G14, respectively.
Where $\RmLS\approx 1$, however, the estimates are definitely off. 

%\Figref{fig:Jcompestr} shows that both $\I_r$ ({\it black dashed %line}) 
%and the simplified expression \ref{eq:JIRE} 
%({\it black dashed-dotted line}) provide comparable estimates   
%which are much better than the radial component 
%of $\sigma \Uv\times\Bv$.  

%\begin{figure}
%\centering
%\includegraphics[draft=false,width=0.75\textwidth]{Figs/%Jr_est_l50l30_profiles_Lucia2_G14_ready.pdf}
%\caption{Radial profiles of the rms radial current
%densities along with three different estimates.}
%\label{fig:Jcompestr}
%\end{figure} 

\begin{figure}
\centering
\includegraphics[draft=false,width=0.75\textwidth]{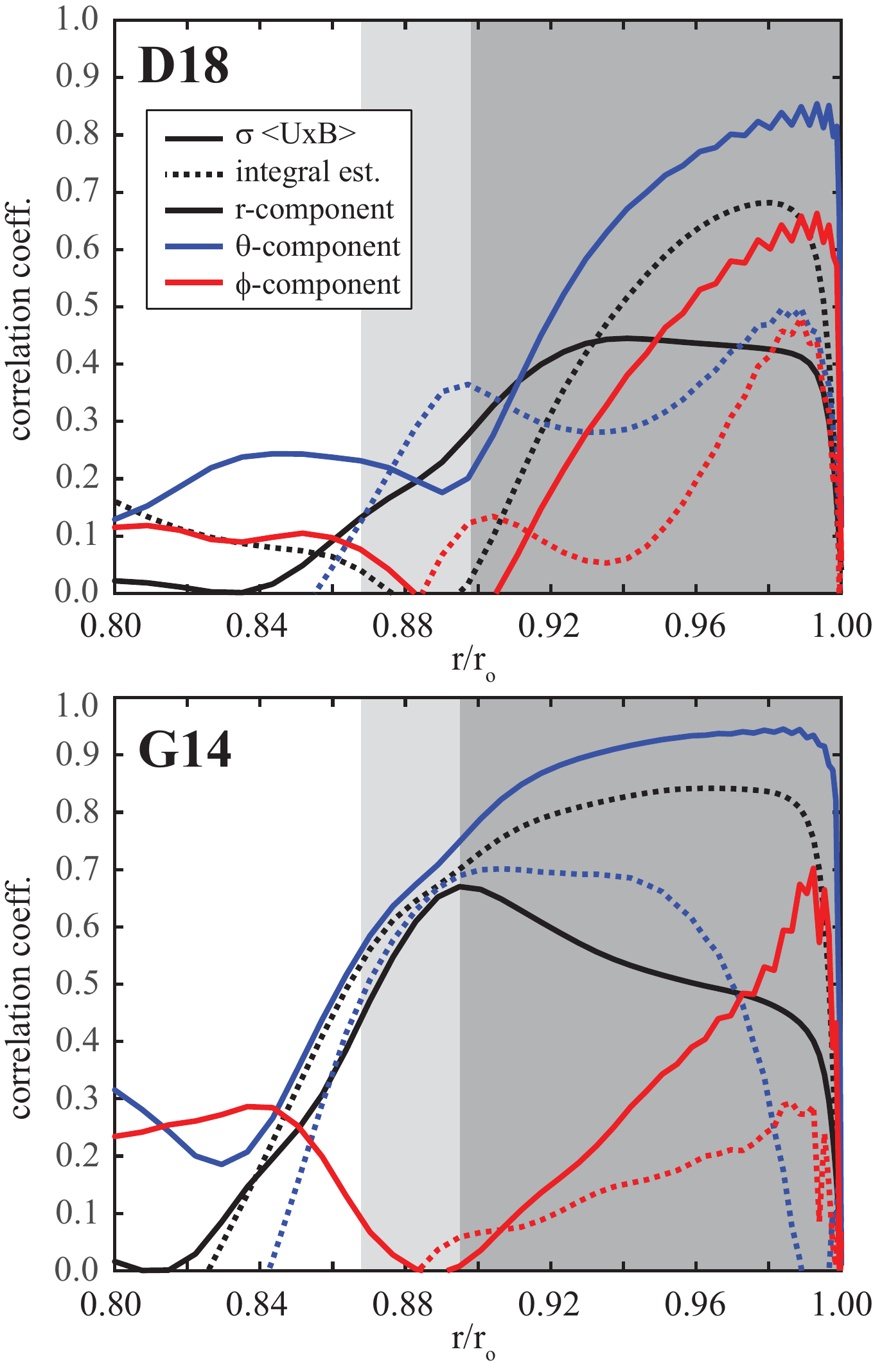}
\caption{Pearson correlation coefficients between 
individual current density components and two different 
estimates. {\it Solid lines} refer to the estimates based on 
the simplified Ohm's law \refp{eq:UxB} while {\it dotted lines} 
refer to the integral expression \refp{eq:UxBI} for the horizontal and 
\refp{eq:JIRE} for the radial component.}
\label{fig:Jcorr}
\end{figure} 

To access whether the estimates not only 
capture the rms amplitudes but also the structure, 
we calculate the Pearson 
correlation coefficients for each radius, for example
\bel{eq:corr}
 C(\J_\theta,\sigma (\Uv\times\Bv)_\theta ) = 
   \frac{\overline{ \left(\J_\theta - \overline{\J}_\theta\right)\;
                    \left( \left[\Uv\times\Bv\right]_\theta - 
             \overline{\left[\Uv\times\Bv\right]}_\theta \right)}}
       {\left[\overline{\left(\J_\theta - \overline{\J}_\theta     
                       \right)^2} \;\;
        \overline{\left(\left[\Uv\times\Bv\right]_\theta -     
            \overline{\left[\Uv\times\Bv\right]}_\theta
                        \right)^2}\right]^{1/2}}
\eec
where the overbars refer to an average over a spherical surface. 
The results, shown in \figref{fig:Jcorr}, demonstrate that 
$\sigma \Uv\times\Bv$ no only provides a better
rms estimate but also a better local estimate than $\Iv_H$ for the 
horizontal current components. 
Particularly high correlations are reached for the latitudinal 
current density in G14 where the zonal flow contributes most strongly.
For the radial component, \eqnref{eq:JIR} also provides a decent 
local estimate with Pearson coefficients up to $0.85$ 
for G14. 

\begin{figure}
\centering
\includegraphics[draft=false,width=0.75\textwidth]{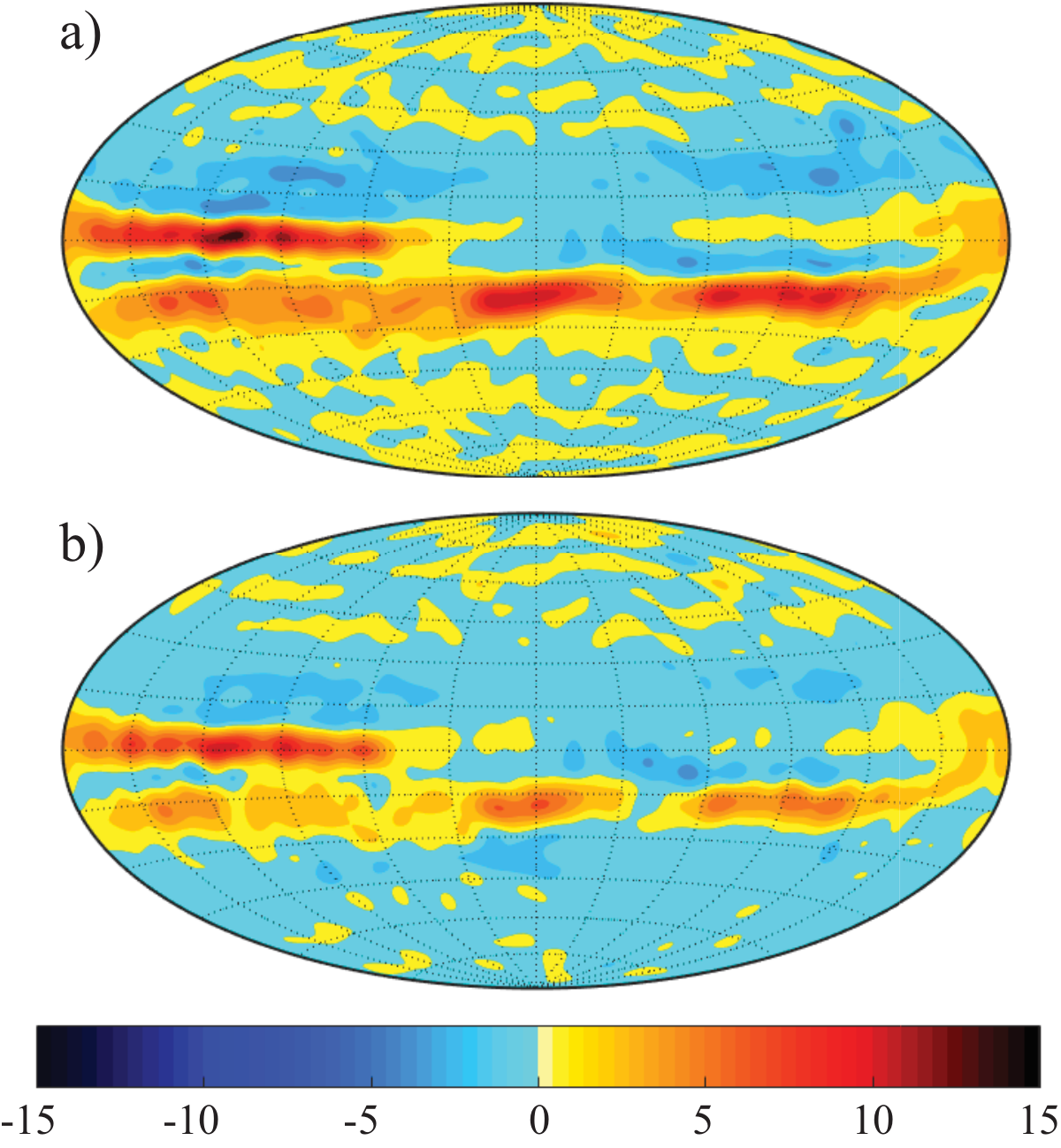}
\caption{Comparison of {\it a)} $\mu J_\theta$ with  
{\it b)} estimate  
$\Pm (\Uv\times\Bv)_\theta$   
at $r=0.94\,r_o$ for the snapshot of dynamo G14.
Dimensionless quantities shown.}
\label{fig:Jtmap}
\end{figure}

\Figref{fig:Jtmap} illustrates the close agreement for the 
latitudinal current density in model G14 at $r=0.94\,r_o$. 
The banded structure due to the zonal flow action already 
reported by \citet{Gastine2014} is clearly apparent. 

\subsection{Estimating the Non-Potential Field}

The toroidal and the non-potential poloidal
fields can be assessed by uncurling 
Ampere's law $\curl \Bv=\mu \jv$. This becomes particularly 
simple in the \SDCR\ where the radial 
field gradients clearly dominate so that   
\bel{eq:Amp}
\uvr\times\frac{\partial}{\partial r}\Bv_H \approx \mu \jv_H
\eep
Integrating this expression from $r_o$ to $r$ yields 
the integral approximation $\Dv_J$, 
\bel{eq:Bint}
\hat{\Bv}_H(r) \approx \Dv_H(r) \:= \mu\;\int_r^{r_o}\,d r^\prime\;\uvr\times\jv_H
\eec
where the integration constant has been set to zero, assuming 
that $\hat{\Bv}_H$ and the current vanish at $r_o$. 

\begin{figure}
\centering
\includegraphics[draft=false,width=0.6\textwidth]{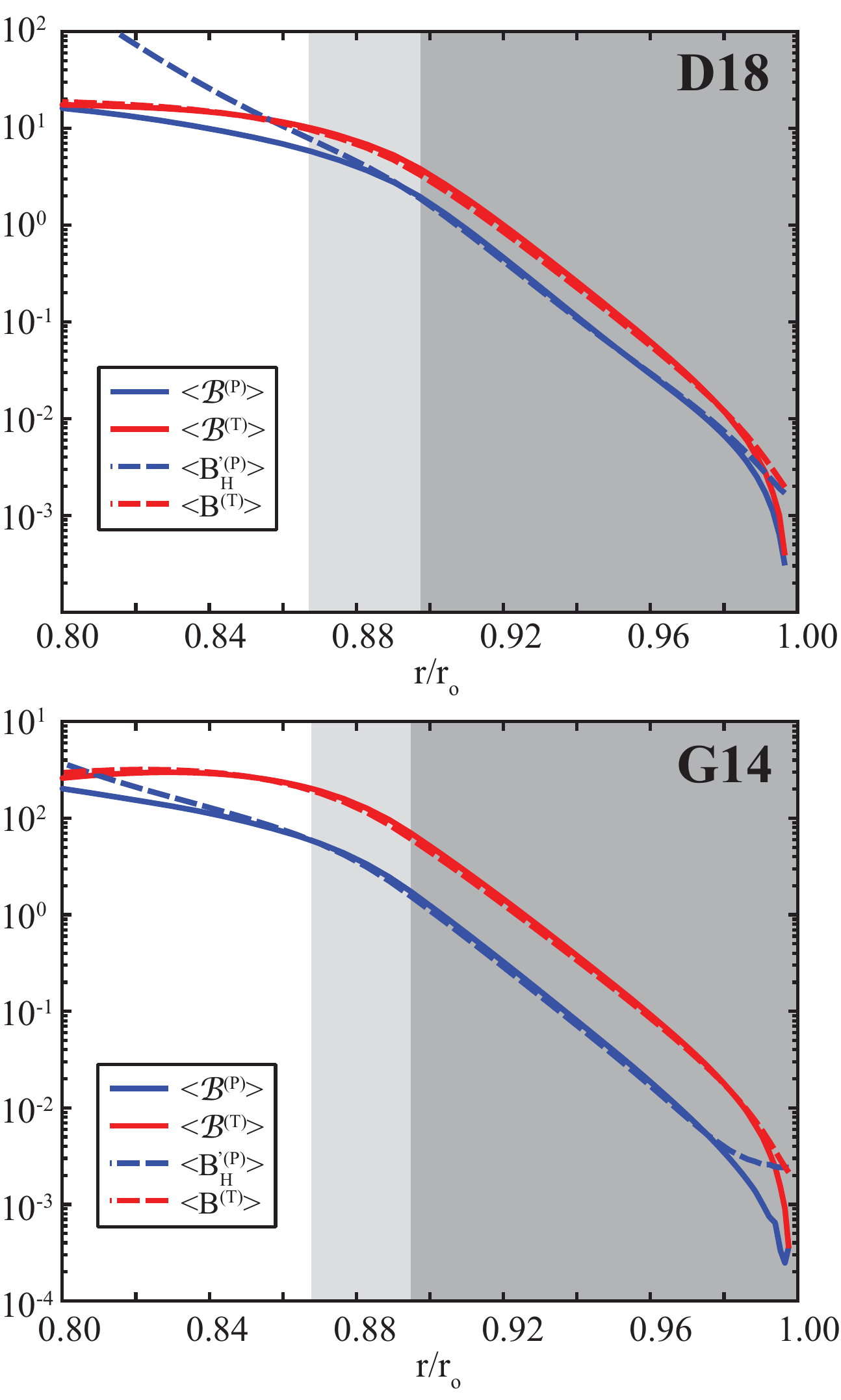}
\caption{Comparison of the rms values for the integral estimates
from uncurling Ampere's law with the rms toroidal field $\langle\Bv^{(T)}\rangle$ and the rms horizontal non-potential field $\langle\hat{\Bv}_H\rangle$. 
Dimensionless quantities
shown.
}
\label{fig:Bnonpot}
\end{figure}

Separating the current into poloidal and toroidal contributions allows for 
individually estimating the toroidal and poloidal non-potential fields. 
Both aggree very well with the horizontal components 
of the toroidal field and the non-potential poloidal field 
$\hat{\Bv}^{(P)}$ in the \SDCR, as is demonstrated in \figref{fig:Bnonpot}. 
The radial integral \refp{eq:Bint} continues to provide a good representation 
of the toroidal field even below the \SDCR. 

Combining these integral expressions with the current approximation via 
the simplified Ohm's law yields the integral estimate $\Dv_O$: 
\bel{eq:BOint}
\hat{\Bv}_H(r) \approx \Dv_O(r) \:= \;\int_r^{r_o}\,d r\;
\;\frac{\uvr\times\left(\Uv\times\Bv\right)}{\lambda}
\eep  
The dominance of the radial dependence of $\lambda$ then 
suggests 
\bel{eq:BUxB}
\hat{\Bv}_H(r) \approx \Dv_O(r) \:= 
\frac{d_\lambda}{\lambda}\;
\uvr\times\left(\Uv\times\Bv\right)
\eep  
and an rms value of 
\bel{eq:Best}
\langle\hat{\Bv}\rangle \approx \RmL \langle\tilde{\Bv}\rangle
\eep

For estimating the poloidal field we rely on the 
respective dynamo equation \eqnref{eq:dtpol}. 
Since radial derivatives clearly dominate, the stationary radial
component of the dynamo equation reduces to  
\bel{eq:dBr}
 \lambda\;\left(\frac{\partial}{\partial r}\right)^2\; \hat{\B}_r \approx 
 - \uvr\cdot\curl \left( \Uv\times\tilde{\Bv} \right)
\eec
where we have used the potential background 
field on the right hand side. 
The diffusive left hand side can reasonably be approximated with
$\lambda \hat{\B}_r / d_\lambda^2$, which yields  
\bel{eq:Brest}
\hat{\B}_r \approx - \frac{d_\lambda^2}{\lambda}\;\uvr\cdot
 \curl \left( \Uv\times\tilde{\Bv} \right) 
\eep

The quality of estimates \eqnref{eq:Best} 
and \eqnref{eq:Brest} will be quantified by the ratios 
\bel{eq:Bratio}
R_{B} = \RmL \frac{\langle\tilde{\B}\rangle}{\langle\hat{\Bv}\rangle}
\ee
and
\bel{eq:BrRatio}
R_{Br} = \frac{ d_\lambda^2}{\lambda} 
\frac{\langle\uvr\cdot\curl \left( \Uv\times\tilde{\Bv} \right) \rangle}
{\langle \hat{\B}_r\rangle}
\eec
respectively.
A simplified expression ignoring the curl and cross product yields 
\bel{eq:Brests}
\langle\hat{\B}_r\rangle \approx  \RmLS \langle\tilde{\Bv}\rangle
\eec
with the respective ratio
\bel{eq:BrRatioS}
\hat{R}_{Br} = 
\RmLS \frac{\langle\tilde{\Bv}\rangle}{\langle\hat{\B}_r\rangle}
\eep

\begin{figure}
\centering
\includegraphics[draft=false,width=0.75\textwidth]{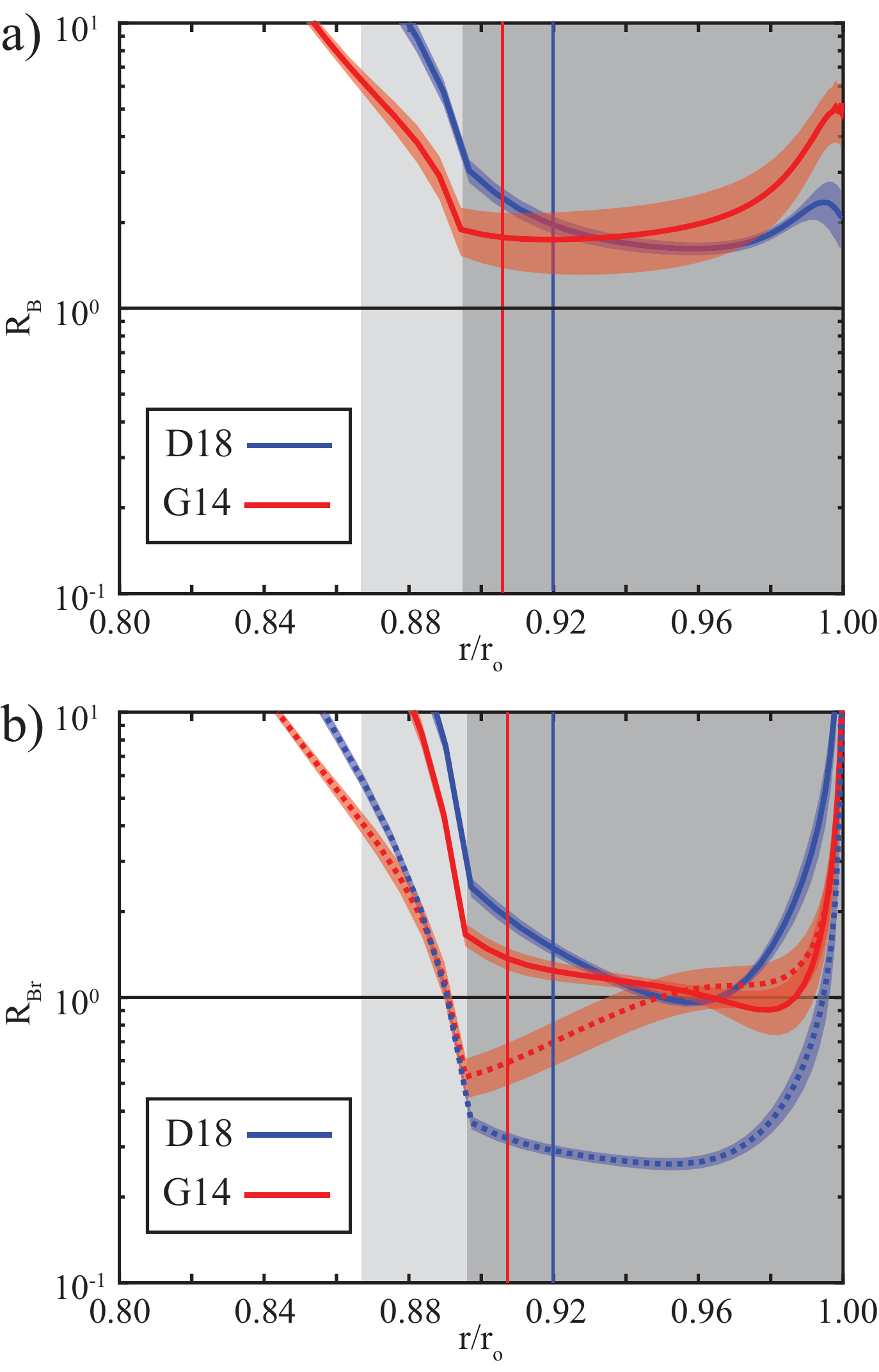}
\caption{Ratio of rms values for the non-potential radial field 
estimate \ref{eq:Brest} and for $\hat{\B}_r$. 
Results have been averaged 
over 14 snapshots for dynamo D18 and 9 snapshots for dynamo G14. 
Transparent background stripes with a width of two standard deviations 
illustrate the variability. The vertical lines mark where 
$\RmL$ ({\it solid}) or $\RmLS$ ({\it dotted}) exceed one. 
}
\label{fig:Bratio}
\end{figure}

\Figref{fig:Bratio} shows that very reasonable values 
between $R_B=1.3$ and $R_B=2$ for the total non-potential field 
and between $R_{Br}=0.8$ and $R_{B_r}=2$ for its radial contributions 
are achieved in the \SDCR, ignoring once more 
the very outer few percent in radius. 
In terms of magnetic Reynolds numbers, the region 
where the estimates offer acceptable results 
extends to a depth between 
$\RmL=1$ and $\RmLS=1$ (see vertical lines in \figref{fig:Bratio}. 

The simplified estimate \refp{eq:BrRatioS}, on the other hand,
underestimates the radial non-potential field by a factor three
for dynamo D18, which has a particularly small scale background field
$\tilde{\Bv}$. For dynamo G14, the simplification has a much 
smaller effect and the quality of the estimate is only
mildly affected. 
%\Figref{fig:Bratio} also shows that once more 
%the estimates remain more or less valid throughout the whole \SDCR.  

\begin{figure}
\centering
\includegraphics[draft=false,width=0.75\textwidth]{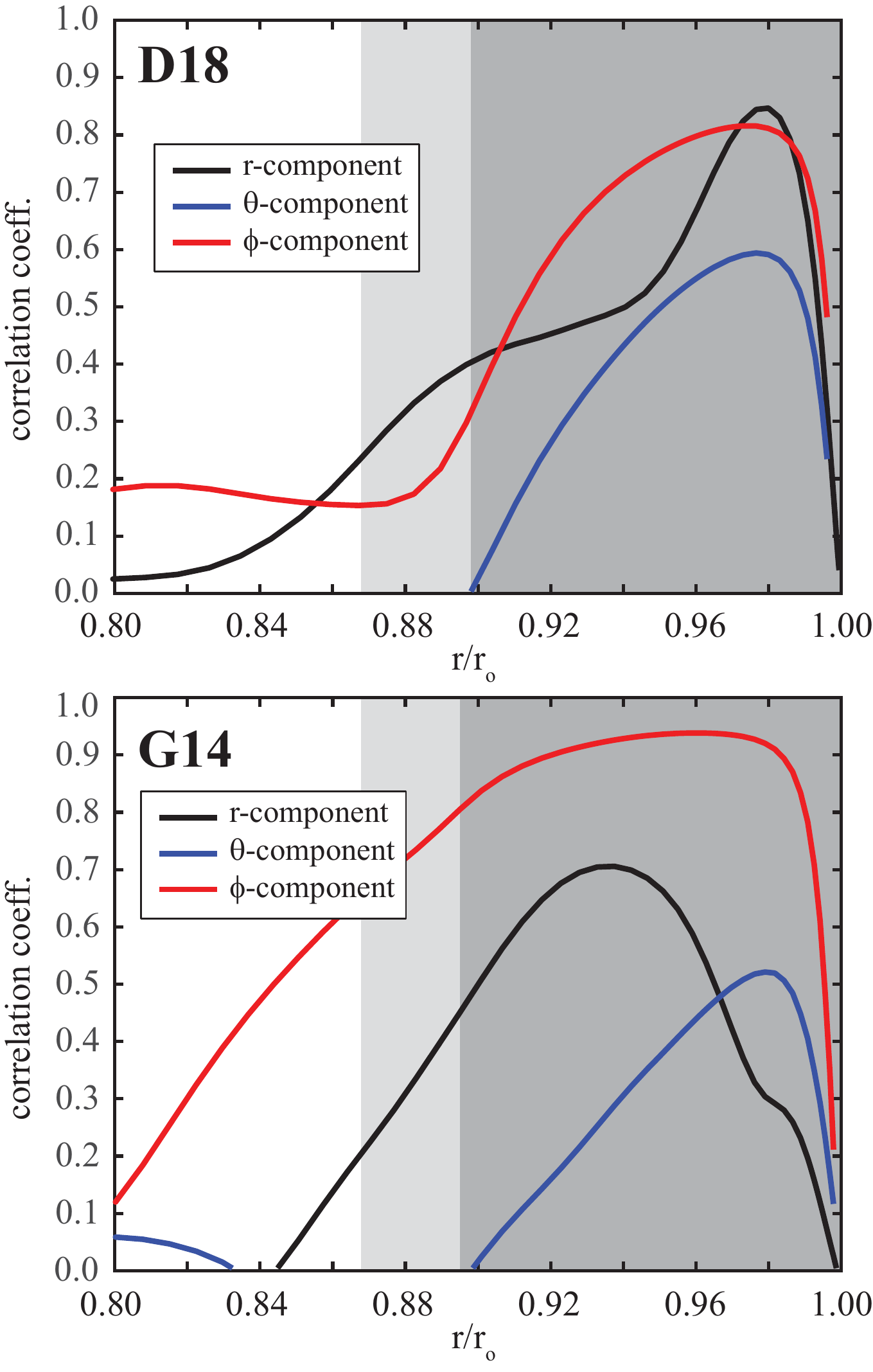}
\caption{Peason correlation coefficients between 
individual magnetic field components and respective estimates. 
}
\label{fig:Bcorr}
\end{figure}

\begin{figure}
\centering
\includegraphics[draft=false,width=0.75\textwidth]{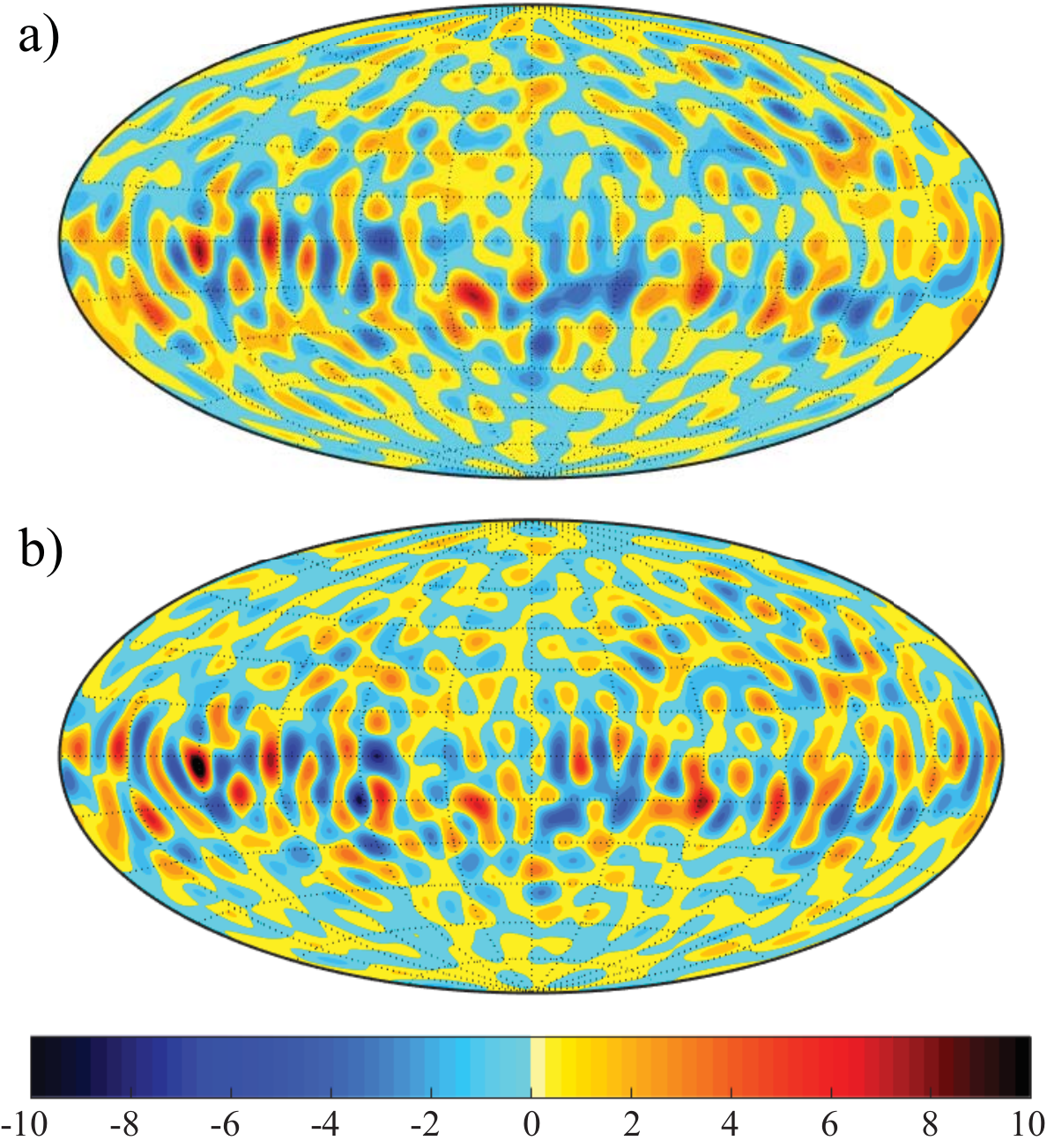}
\caption{Comparison of {\it a)} $\hat{\B}_r$ with  
{\it b)} estimate \eqnref{eq:Brest}  
for $r=0.94 r_o$ and a snapshot for dynamo G14. 
Dimensionless quantities
shown.
}
\label{fig:BrMap}
\end{figure}

The Pearson correlation coefficients in the \SDCR\ 
range between $0.5$ and $0.9$ for 
the non-potential radial field estimate, 
and between $0.6$ and $0.8$ for 
the non-potential azimuthal field estimate, 
as is illustrated in \figref{fig:Bcorr}. 
For the weak latitudinal field, however, the estimate is much
less reliable. 
\Figref{fig:BrMap} illustrates the good agreement between $\hat{\B}_r$ 
and estimate \refp{eq:Brest} for the G14 snapshot.  
The banded structure becomes once more apparent, but the field 
is much more small scale owed to the complex convective 
flows that provide the main radial field induction \citep{Gastine2014}.

\subsection{Estimating the Lorentz Force}

\begin{figure}
\centering
\includegraphics[draft=false,width=0.75\textwidth]{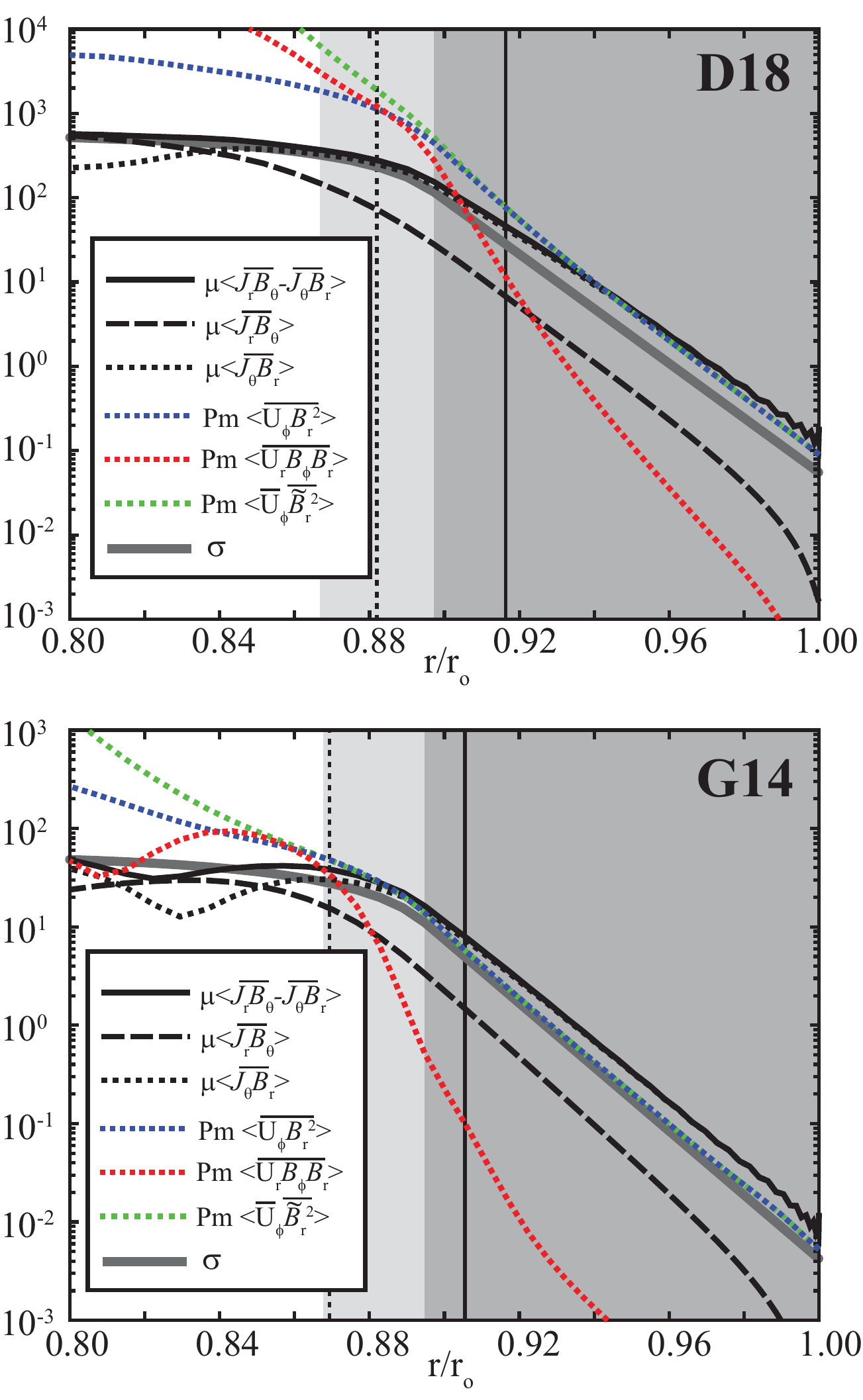}
\caption{Zonal rms Lorentz force contributions and estimates.
See text for more explanation. 
Dimensionless quantities shown.}
\label{fig:LF}
\end{figure}

Having derived estimates for electric currents and magnetic fields, 
we can combine both to also assess the Lorentz force $\Lv$ 
in the \SDCR. 
Of particular interest is its zonal (axisymmetric azimuthal) component  
$\overline{L}_\phi$ which could potentially impact the zonal winds. 
\Figref{fig:LF} shows the profiles of the rms contributions to 
$\langle\overline{L}_\phi\rangle$ for dynamos D18 and G14. 
The figure illustrates that from the two contributions,
\bel{eq:LFA}
\overline{L}_\phi = \overline{J_r B_\theta} - \overline{J_\theta B_r}
\eec
the second clearly dominates because of the 
higher latitudinal current density. 
%And not surprisingly, the profile of the Lorentz force closely
%follows the electrical conductivity profile. 

\begin{figure}
\centering
\includegraphics[draft=false,width=0.75\textwidth]{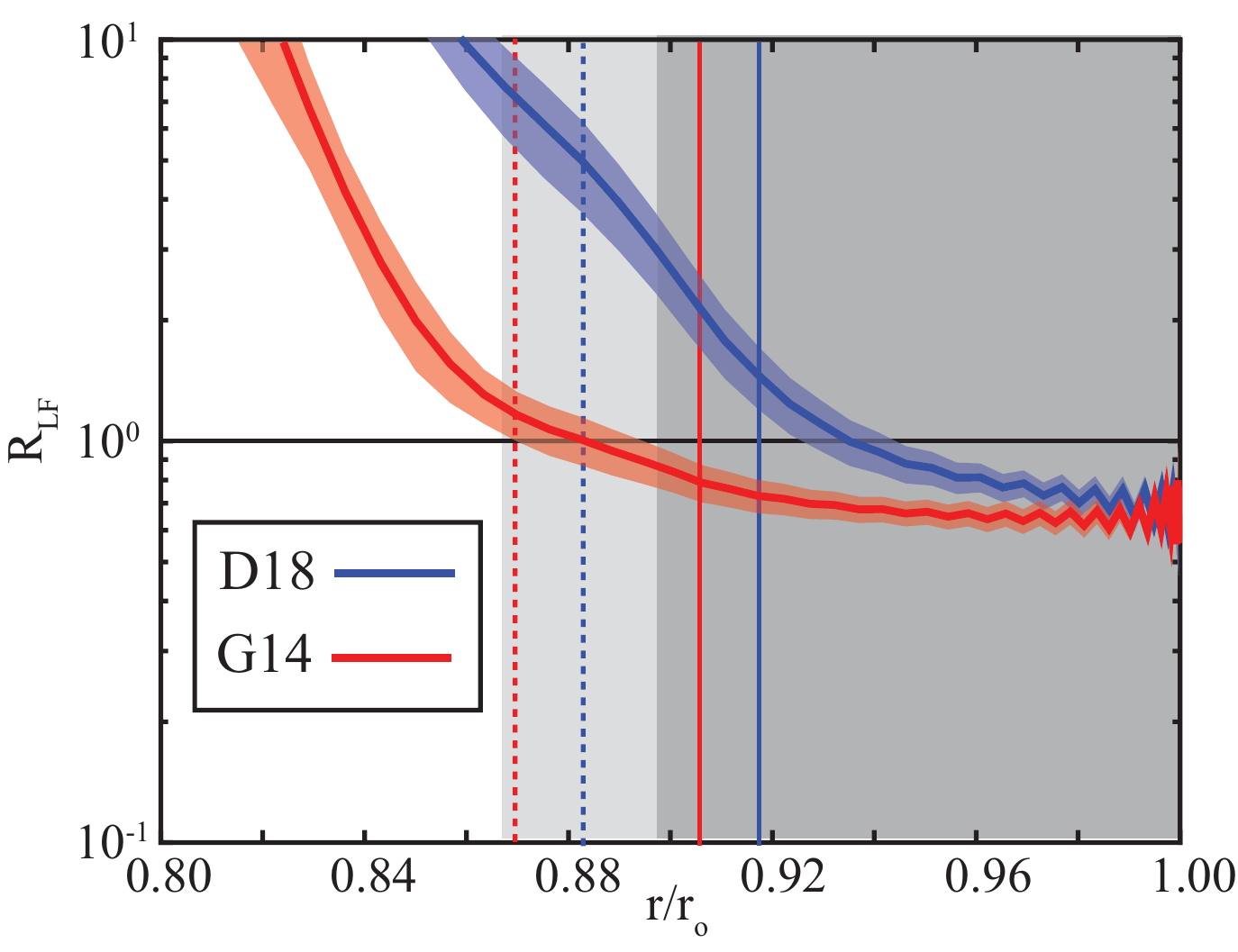}
\caption{Ratio of estimate \refp{eq:LFAZ} to the 
zonal Lorentz force. {\it Thick lines} show
averages over 14 snapshots for dynamo D18 ({\it blue}) and 10 snapshots 
for dynamo G14 ({\it red}). The transparent background stripes with a 
width of twice the standard deviation illustrate the variability.
Thin vertical lines show where 
$\RmL$ ({\it solid}) or $\RmLS$ ({\it dotted}) exceed one.}
\label{fig:LFratio}
\end{figure}

Combining the estimates for $J_\theta$ and $B_r$ then suggest 
\bel{eq:LFE}
\overline{L}_\phi \approx - \sigma\;
\overline{\left(\Uv\times\Bv\right)_\theta  \B_r } = 
-\sigma\;\overline{\U_\phi\B_r^2 } +\sigma\;\overline{\U_r\B_\phi\B_r}
\eep
\Figref{fig:LF} illustrates that the first contribution dominates, 
in particular for model G14 where zonal flows are stronger. 
We can thus estimate the rms zonal Lorentz force based on a conductivity 
model, a zonal flow model, and the potential field:   
\bel{eq:LFAZ}
\langle \overline{L}_\phi \rangle 
\approx \sigma \;
\langle \overline{\U}_\phi \rangle\;\langle\overline{\tilde{\B}^2_r}\rangle 
\eec
The respective ratio 
\bel{eq:LFratio}
R_{LF} = \frac{\sigma\;\langle\overline{\U}_\phi\rangle\;
\langle\overline{\tilde{\B}_r^2}\rangle}
{\langle \overline{L}_\phi\rangle}
\eec
shown in \figref{fig:LFratio}, demonstrates that this 
expression tends to underestimate the
zonal Lorentz force by only up to $40$\%. 
The region where this estimates can reasonable 
be applied is larger for G14 where zonal flows are stronger. 

%Estimate \refp{eq:LFAZ} suggests that the ratio of 
%Lorentz to Coriolis force is
%given by the depth dependent Elsasser number:
%\bel{eq:Els}
% \Lambda = \frac{\sigma \langle\tilde{\B}\rangle^2}{2\rho\Omega}
%\eep
%While an Elsasser number of often not be a good proxy for the relative 
%importance of the Lorentz force in a typical dynamo 
%region \citep{Wicht2010,Soderlund2015} it 
%seems to do a decent job in the \SDCR\ where the 
%dynamo dynamics is quasi stationary.
\section{Discussion and Conclusion}
\label{sec:Discussion}

The analysis of our numerical simulations shows that dynamo action in the Steeply Decaying Conductivity Region (\SDCR) is dominated by Ohmic dissipation.
The magnetic field dynamics becomes quasi stationary with a good balance between induction and diffusion. Ohm's law, on the other hand, assumes the simplified 
form for a fast moving conductor, where the electric field can be neglected. 
Electric currents and the toroidal field simply decay 
with the electrical conductivity, while the poloidal 
field approaches a potential field.  

This particular situation allows to formulate rather simple 
estimates based on the knowledge of the surface field, 
the conductivity profile, and the flow. 
The electric current density can 
be estimated via the simplified Ohm’s law with 
a suggested  rms value of 
$\langle\Jv\rangle \approx \sigma \langle\Uv\rangle\;\langle\tilde{\Bv}\rangle$, 
where $\tilde{\Bv}$ is the downward continued surface field under 
the potential field assumption. 
The accuracy is higher than the more classical estimate
used by \citet{Liu2008} for their assessment of Ohmic heating 
due to Jupiter's zonal winds. 

Also of interest is the locally induced radial magnetic field
that could potentially be detected by the Juno spacecraft. 
Our analysis shows that 
$\langle\B^\prime_r\rangle\approx\RmLS \langle\tilde{\Bv}\rangle$ 
provides a reasonable estimate, where $\RmLS$ is a modified 
magnetic Reynolds number that depends on the square of the 
diffusivity scale height $d_\lambda$. 
The locally induced toroidal field, on the other hand,
can be predicted via 
$\langle\Bv\rangle \approx \RmL \langle\tilde{\Bv}\rangle$ 
and is thus by a factor $ \RmL / \RmLS=d / d_\lambda$ larger than
the radial field. 
When using the conductivity model by \citet{French2012}, 
this predicts that the locally induced toroidal field 
is about $10^3$ time larger than the locally induced radial
field at $0.96\,r_J$. At $0.9\,r_J$, this ratio has decreased 
to $10^2$.  

While the toroidal field estimate agrees with the 
assessment by \citet{Cao2017}, the poloidal field estimates differ. 
\citet{Cao2017} consider
the poloidal field produced by non-axisymmetric (helical) flows 
acting on the local toroidal field, but the 
advective modification of $\tilde{\Bv}$ turns
out to be significantly larger in the \SDCR\ of our simulations. 

\citet{Duarte2018} define the top of the dynamo region
$r_D$ based on a critical magnetic Reynolds number.  
A more reasonable local definition is the depth where the 
local magnetic effects reach a certain threshold. 
Since the electric current, toroidal field, and radial field 
all scale with different magnetic Reynolds numbers, the answer
will actually depend on the quantity considered. 

Choosing the radial magnetic field has the advantages that it is 
most directly connected to magnetic field observations. 
For the most recent field models based on Juno data, 
\citet{Connerney2018} discuss an $r_D$ value of $0.85\,r_J$. 
The authors argue that the spectrum of $\tilde{\Bv}$
is close to a white spectrum at this depth, a criterion that 
successfully predicts Earth's core-mantle boundary 
when excluding the axial 
dipole contribution. 
However, $0.85\,r_J$ lies below the transition to the 
metallic Hydrogen region \citep{French2012}
and below the anticipated depth of the zonal winds \citep{Kaspi2018}. 
Assuming a mean convective velocity of $3\,$cm/s \citep{Gastine2014} 
and the French conductivity profile predicts 
$\RmL=2.7\times 10^4$ and $\Rm=7.6\times 10^5$ at $0.85\,r_J$. 
These huge values suggest that dynamo action should already 
be more than well developed. 

The respective $\RmL$ profile actually exceeds unity at 
about $0.93\,r_J$, which will therefore 
roughly mark the depth the estimates discussed here 
start to loose their basis. 
Since $\RmLS$ is about $10^{-4}$ at $0.93\,r_J$ 
and the potential field amplitude is about $1\,$mT, 
the locally induced radial field would be roughly 
$10^{-4}\,$mT. 

However, these considerations ignore the zonal winds, 
which reach amplitudes of around $150\,$m/s at Jupiter's cloud level. 
Recent consideration based on Juno's gravity measurements 
constrain the depth profile of the zonal winds and 
suggest a considerably lower wind speed at about $0.96\,r_J$. 
In a forthcoming paper, we use the estimates derived here 
in combination with the new depth profiles for the zonal winds 
to predict the zonal flow induced magnetic fields 
and electric currents and also calculate the related Ohmic heating. 
The analysis presented here suggest that decent local estimates 
are also possible. The predicted maps of Ohmic heating 
and radial field modifications could be compared with 
spacecraft observation to detect 
potential zonal field induction effects.  

The questions addressed here are also of interest for other planets 
where fast fluid flows correlate with a region of 
decaying electrical conductivity. Saturn, Uranus, and Neptune, 
come to mind. The Ohmic heating in the outer atmospheres of hot-Jupiters, where
the ionization of alkali metals creates a \SDCR, is a possible mechanism to explain the particularly low density (inflation) of some of these exoplanets. \citet{Batygin2010} 
assume a stationary dynamo equation to estimate that the locally 
induced currents would indeed provide sufficient heating power. 
Our result suggest that their approach and thus also likely their
conclusions are viable. 

%Bibliography:
\clearpage
\bibliography{JWbib}
\bibliographystyle{elsart-harv}

\renewcommand{\abstractname}{Acknowledgements}
\begin{abstract}
This work was supported by the German Research Foundation (DFG) in the 
framework of the special priority program PlanetMag (SPP 1488). 
The MHD code MagIC used for the simulations is freely available 
on GitHub: https://github.com/magic-sph/magic. The simulation
data for the two cases expored here and the Matlab package 
used for the analysis can be downloaded from https://dx.doi.org/10.17617/3.1q. 
\end{abstract}

\end{document}